\documentclass[final,3p,times]{elsarticle}
\usepackage{amssymb}
\usepackage{threeparttable}

\usepackage{xcolor}
\definecolor{blue}{RGB}{0,0,0}    % Medium blue
\usepackage{subcaption} % Ensure this is in your preamble

\usepackage{amsmath}
\usepackage{colortbl} % For coloring
\usepackage{xcolor} % For text coloring
\usepackage{tcolorbox}
\usepackage{threeparttable}
\usepackage{url}

\journal{Journal of Systems and Software}

\begin{document}

\begin{frontmatter}



\title{
From Attack Descriptions to Vulnerabilities: A Sentence Transformer-Based Approach
}


% \author{Refat, Bruno, Barbara} %% Author name

%% Author and Affiliation 1
\author[label1]{Refat Othman} 
%% Affiliation 1
\affiliation[label1]{organization={Free University of Bozen-Bolzano},
            % addressline={NOI Techpark},
             city={Bolzano},
             postcode={39100},
             country={Italy}
             }

%% Author and Affiliation 2

\author[label1]{Diaeddin Rimawi}
%% Affiliation 2

\author[label2]{Bruno Rossi}
%% Affiliation 2
\affiliation[label2]{organization={Masaryk University},
             city={Brno},
             postcode={60200},
             country={Czech Republic}
             }

%% Author and Affiliation 3
\author[label1]{Barbara Russo}

%% Abstract
\begin{abstract}
In the domain of security, vulnerabilities frequently remain undetected even after their exploitation.
In this work, vulnerabilities refer to publicly disclosed flaws documented in Common Vulnerabilities and Exposures (CVE) reports. 
Establishing a connection between attacks and vulnerabilities is essential for enabling timely incident response, as it provides defenders with immediate, actionable insights. However, manually mapping attacks to CVEs is infeasible, thereby motivating the need for automation.
% Establishing a connection between attacks and vulnerabilities is crucial as it aids security professionals in promptly identifying and addressing security incidents.
This paper evaluates 14 state-of-the-art (SOTA) sentence transformers for automatically identifying vulnerabilities from textual descriptions of attacks. 
Our results demonstrate that the \texttt{multi-qa-mpnet-base-dot-v1 (MMPNet)} model achieves superior classification performance when using attack Technique descriptions, with an F$_1$-score of 89.0, precision of 84.0, and recall of 94.7.
Furthermore, it was observed that, on average, 56\% of the vulnerabilities identified by the \texttt{MMPNet} model are also represented within the CVE repository in conjunction with an attack, while 61\% of the vulnerabilities detected by the model correspond to those cataloged in the CVE repository.
A manual inspection of the results revealed the existence of 275 predicted links that were not documented in the MITRE repositories. Consequently, the automation of linking attack techniques to vulnerabilities not only enhances the detection and response capabilities related to software security incidents but also diminishes the duration during which vulnerabilities remain exploitable, thereby contributing to the development of more secure systems.
\end{abstract}


% \begin{highlights}
% \end{highlights}

%% Keywords
\begin{keyword}


Cyber Threat Intelligence \sep MITRE ATT\&CK \sep CAPEC \sep CVE \sep Sentence Transformer\sep Attack-vulnerability linking 
\end{keyword}

\end{frontmatter}

\section{Introduction}
\label{sec:introduction}
%Problem once an attack manifests itsself we need to be fast in detecting and mitigating the vulnerablities exploited throuhg the attack. 

%The time in which  vulnerabilities are not linked to the attack or not soon linked to the attack, the system remains expose to malicious activities. This costs.

Cybersecurity represents a fundamental challenge in the protection of modern systems, which are continually exposed to evolving and complex cyber threats~\cite{admass2024cyber}. Once a cyberattack manifests, immediate action is essential to detect and mitigate the vulnerabilities exploited during an incident. When these vulnerabilities are not quickly identified and linked to the attack, systems remain exposed to malicious activities, resulting in substantial financial and operational losses. In fact,
cyberattacks are expected to cost organizations \$3 trillion in 2015, \$6 trillion in 2021, and more than \$10.5 trillion annually by 2025~\cite{Cybercrime}. Moreover, an attacker targets an organization's system over a thousand times a week on average~\cite{CResearch}. 
% The attackers have enough expertise and an extensive spectrum of malicious talents. 
However, cybersecurity professionals can leverage Cyber Threat Intelligence (CTI) to proactively defend against cyberattacks~\cite{rahman2023attackers}.
In this context, well-structured cybersecurity knowledge repositories are crucial for effective defense. These resources assist security analysts in understanding and tracing connections between a system's vulnerabilities and the methods used to exploit them.
% between how systems are attacked and where they are vulnerable. 
% In this context, it is necessary to use cybersecurity knowledge databases and cyber information resources to identify and control continuously developing cyber risks. 
%exiss knowldge about Attcks and vulnerabilities

The MITRE Corporation has created a collection of resources, such as Adversarial Tactics, Techniques, and Common Knowledge (ATT\&CK)~\cite{ATTACK}, Common Attack Pattern Enumeration and Classification (CAPEC)~\cite{CAPEC}, Common Weakness Enumeration (CWE)~\cite{CWE}, and Common Vulnerabilities and Exposures (CVE)~\cite{CVEdataset}. ATT\&CK provides tactics, techniques, and procedures (TTPs) used by cyber attackers. CAPEC represents a publicly available catalog of common attack patterns helping to understand how attackers can exploit weaknesses~\cite{CAPEC}.
% is called an Attack Pattern, which is \textit{``the common approach and attributes related to exploiting a weakness in a software, firmware, hardware, or service component''}
 In addition, CAPEC is structured as a list of information about the attack, including techniques for the attack, potential outcomes, and mitigations~\cite{refat2024comparison}.
CWE is a community-developed collection of typical weaknesses in software, coding errors, and security flaws. 
CVE is a standardized dictionary of common terms for publicly known cybersecurity vulnerabilities. All these initiatives aim to enhance the efficiency of identifying, finding, and fixing vulnerabilities by providing a unified naming system~\cite{WhatCVE, othman2024vulnerability}. The CVE entries are widely used, but CVE issues often lack contextual information about the attack techniques that may exploit them~\cite{sonmez2021classifying,elder2022really}.

%Navigating and find the attacks or linking them to vulnerabilities may be a too long activity with not always some useful results . Autmoating the process is amusty

% MITRE repositories such as ATT\&CK, CAPEC, CWE, and CVE provide valuable and structured knowledge. However, they are maintained independently and often lack cross-references between related entities. For example, many CVE reports contain rich vulnerability descriptions, but they rarely reference which attack techniques might exploit them. Similarly, ATT\&CK entries outline adversary behaviors without explicitly citing the CVEs they could trigger. Thus, navigating MITRE repositories to identify links between attacks and vulnerabilities is often a time-consuming and inefficient task. 
MITRE repositories such as ATT\&CK, CAPEC, CWE, and CVE provide valuable and structured knowledge. However, because these repositories are normally maintained manually and evolve independently, they often lack cross-references between related entities. 
For example, many CVE reports contain detailed vulnerability descriptions but rarely indicate which attack techniques might exploit them, while ATT\&CK entries describe adversary behaviors without explicitly citing the CVEs they could trigger. Consequently, analysts who need to trace from an observed technique to actionable CVE identifiers encounter a fragmented search process that is time-consuming and prone to omissions.
% Manual efforts are not only slow but also prone to inconsistencies, particularly given the complexity and rapidly evolving nature of cyber threats. 
% Moreover, vulnerabilities frequently remain undetected even after exploitation, enabling attackers to extend their presence within targeted systems. In fast-evolving attack scenarios, delays in linking attacks with the exploited vulnerabilities can delay effective response and remediation efforts. Thus, linking ATT\&CK techniques and CVE vulnerabilities could offer immediate, actionable insights for defenders. However, manually mapping 625 ATT\&CK techniques to more than 295,000 CVE entries~\cite{CVEdataset} is infeasible. Automating this mapping process not only accelerates threat mitigation but also supports timely decision-making and strengthens overall defensive mechanisms.

Given the scale of these resources, with 625 ATT\&CK Techniques and more than 295{,}000 CVEs~\cite{CVEdataset}, manual mapping is infeasible. This burden is compounded by the complexity and rapid evolution of cyber threats, which introduce latency and inconsistencies in manual curation and leave essential links missing or outdated. As a result, vulnerabilities may remain undetected even after exploitation, and delays in linking attacks to the exploited vulnerabilities hinder timely response and remediation. Systematically linking ATT\&CK techniques to CVE vulnerabilities would provide immediate, actionable insights for decision-makers. Therefore, automating this process is crucial, as it accelerates threat mitigation, supports timely decision-making, and strengthens defensive mechanisms.
% 
% Navigating these repositories and finding the links between attacks and vulnerabilities is often a time-consuming and inefficient task.} Thus, linking ATT\&CK to CVE issues will help the cyber community, as these two repositories are currently separated. However, linking 625 Attack Techniques manually with 295,604 CVE issues ~\cite{CVEdataset} is a non-trivial task. 
% Therefore, automating the linking between attack descriptions and known vulnerabilities proves highly beneficial, as it accelerates mitigation efforts, supports more informed decision-making, and strengthens the overall effectiveness of defensive security mechanisms. 
% Moreover, identifying which vulnerabilities are frequently triggered by specific tactics, techniques, or procedures provides organizations with actionable insights for risk prioritization and the development of proactive defense planning.}

%Our solution leverages transformer-based sentence embeddings augmented with a similarity layer to predict vulnerabilities from an attack text. Expain

A key insight underlying our work is that when an attack description and a vulnerability report share semantically similar characteristics, such as affected software components, exploited behaviors, or observable consequences, then it is possible to infer a context link between them using sentence transformers. 
 For instance,  ATT\&CK Technique T1566.001 (Spearphishing Attachment) involves adversaries delivering emails with malicious file attachments to deceive users and execute payloads. This behavior semantically aligns with several CVEs that exploit similar deception techniques, 
including spoofed file types, misleading download dialogs, and inconsistently rendered file names (e.g., \texttt{CVE-2001-0398}, \texttt{CVE-2002-0722}, \texttt{CVE-2004-1104}, \texttt{CVE-2005-1575}, and \texttt{CVE-2005-0243}). These semantic overlaps suggest that the link between attack techniques and vulnerabilities can be inferred by analyzing the linguistic similarity between their textual descriptions. Thus, our solution leverages transformer-based sentence models,
which are pre-trained to understand semantic relationships between textual inputs. These models encode both attack and vulnerability descriptions into fixed-size embeddings in a shared vector space. Computing cosine similarity between these embeddings to predict the links between attack and vulnerability descriptions can help define how closely these embeddings relate in the vector space, allowing for the discovery of meaningful links that can be missed manually. To address this challenge, our previous work introduced VULDAT~\cite{othman2024cybersecurity}, a tool for linking MITRE ATT\&CK techniques to CVE vulnerabilities using MPNet-based sentence embeddings. While effective, VULDAT focused only on Technique descriptions and evaluated nine transformer models.

In this paper, we propose a more comprehensive solution that extends VULDAT by evaluating 14 state-of-the-art (SOTA) transformer models for automated vulnerability detection from cyberattack text.  Our approach leverages the SOTA sentence transformers to link the textual description of an attack (Tactic, Technique, Procedure, and Attack Pattern) to the textual description of CVE reports, demonstrating the model’s effectiveness in both validating known links and discovering new links in the MITRE repositories.
Through automation, our approach reduces the time and effort required for vulnerability identification and provides timely, actionable data that enables cybersecurity professionals to respond to threats more quickly and effectively.
% In this paper, we extend that work by proposing a more comprehensive approach for automated vulnerability detection from cyberattack text. Our approach leverages 14 SOTA sentence transformers to link the textual description of an attack (Tactic, Technique, Procedure, and Attack Pattern) to the textual description of CVE reports. 
To validate our approach, we have built an annotated dataset with explicit links found in MITRE repositories. Our approach's dataset and source code are available on GitHub~\cite{VULDAT}. 
%Why our solution solves the problem
Thus, we intend to answer the following research questions:
\par\noindent
\newcommand{\rqone}{\textbf{RQ$_1$:} \textit{(Transformers' comparison) Which combination of attack information and sentence transformer model achieves the best performance in detecting vulnerabilities?}} 
\newcommand{\rqtwo}{\textbf{RQ$_2$:} \textit{(Best models' effectiveness) To what extent do the best transformer models correctly detect vulnerabilities from attack descriptions?}}
 \newcommand{\rqthree}{\textbf{RQ$_3$:} \textit{(Recommending CVEs) To what extent can our approach recommend missing links between attack techniques and vulnerabilities?} }
\begin{itemize}
    \item \rqone
    
    To answer this question, we evaluate 14 SOTA sentence transformer models, comparing their performance in vulnerability prediction. These models, listed in Table~\ref{tab:pretrainedmodels}, include architectures like Bidirectional Encoder Representations from Transformers (BERT)~\cite{reimers2019sentence} and Text-to-Text Transfer Transformer (T5)~\cite{raffel2020exploring}, optimized for tasks such as semantic search and text classification.
     The goal is to find the most effective sentence transformer model for identifying attack-related textual descriptions of vulnerabilities.
    % \item \textbf{RQ$_2$: }(ATT\&CK descriptions)\textit{ What kind of attack description can be used to accurately predict software vulnerabilities?}
    In addition to comparing model performance, we analyze how different types of attack descriptions influence the performance of the sentence transformers. Specifically, we examine four types of attack information derived from the MITRE ATT\&CK and CAPEC repositories: Tactic, Technique, Procedure, and Attack Pattern. 
    \textcolor{blue}{We hypothesize that Technique descriptions are the most informative and will yield superior results, while larger or semantically optimized transformer models are expected to outperform lightweight models due to their richer contextual embedding.}
    By evaluating each model on these distinct information types, we aim to determine which model performs best overall and which kind of attack description most effectively supports vulnerability detection.
 \item \rqtwo
 
    To answer this question, we evaluated the overlap between the explicit links from the ground truth dataset and our detection list of CVE issues generated by our approach. Our goal was to assess how accurately the model can reproduce existing links and whether it can identify additional, potentially missing ones. 
     \textcolor{blue}{We hypothesize that the best-performing models will achieve a substantial but incomplete overlap with the ground truth, as sentence transformers are expected to capture most of the documented CVE associations while also producing new links due to the inherent incompleteness of the MITRE repositories.}
     
    To this aim, we measured performance using three metrics: Jaccard Similarity, Mapping Accuracy, and Detection Accuracy. The aim is to show our approach's in detecting CVE issues related to the textual description of attacks. 
   \item \rqthree
% To address this question, we investigated whether our approach can discover new links between attacks and vulnerabilities that are not explicitly available in the MITRE repositories.

To address this question, we examined whether our approach can identify potential CVE links that are not explicitly available in the MITRE repositories. By analyzing cases where the model predicts CVEs not found in the ground truth, we assessed the relevance of these "missing" links. 
\textcolor{blue}{We hypothesize that our approach can uncover undocumented but semantically valid attacks to vulnerability links, indicating that MITRE repositories are incomplete and can be enriched through automated linking.}
Our goal is to evaluate the approach’s ability to recommend semantically valid but undocumented connections that could enrich existing vulnerability databases and support proactive threat analysis.

\end{itemize}
Overall, the major contributions of our work are the following:
\begin{itemize}
    % \item The proof-of-concept approach that supports CTI research by providing an  automated way of detecting software vulnerability descriptions and weaknesses from attack reports;
    \item \textcolor{blue}{A proof-of-concept approach that supports CTI research by automatically linking vulnerability descriptions from attack reports;}
    \item \textcolor{blue}{An automated approach for linking attacks to vulnerabilities, validated through an evaluation of 14 SOTA sentence transformers;}
    % \item A comparative analysis on 14 SOTA sentence transformers for vulnerabilities description detection;
    \item A novel annotated mapping dataset~\cite{VULDATDataSet} explicitly linking ATT\&CK with vulnerabilities found in MITRE repositories;
    \item  Empirical analysis of the impact of attack description types in vulnerability detection performance with 275 new validated links between known attacks and vulnerabilities;
\end{itemize}

The paper is structured as follows: Section 2 briefly summarises the background, key concepts, and prior studies on vulnerabilities. Section 3 discusses the methodology, including our approach and dataset. Section 4 shows the results. Section 5 identifies the limitations and summarizes the threats to validity. Section 6 discusses the related work, and the paper concludes in Section 7.

\section{Background}
\label{ch:keyconcepts}
In this section, we review the key concepts relevant to our study: vulnerabilities, weaknesses, as well as attacks and the knowledge bases that have been used in this study. We also describe the sentence transformer models as they have been used extensively in the current work.
\subsection{Vulnerability knowledge bases}
\label{sec:vulnerability}
% We gather information on vulnerabilities from the CVE and CWE repositories.
There are several definitions for vulnerabilities. A \textbf{vulnerability} is defined as a flaw in software, firmware, hardware, or a service component resulting from a weakness that can be exploited, causing a negative impact on the confidentiality, integrity, or availability of an impacted component or components~\cite{dong2023dekedver,elder2024survey}. NIST defines a \textbf{software vulnerability} as "a security flaw, glitch, or weakness found in software code that could be exploited by an attacker (threat source)"~\cite{dempsey2017automation}. A \textbf{ weakness} is a condition in the software, firmware, hardware, or service components that, under certain circumstances, could contribute to introducing vulnerabilities~\cite{CWE, esposito2024validate}. Weaknesses are related to vulnerabilities -- in the sense that they refer to problems that can reduce the security of a system, even if no actual exploit has been identified. When an attacker finds a way to exploit a weakness, then it becomes a vulnerability. 
 For instance, through a weakness, an attacker might be able to assume the identity of a superuser or system administrator and get complete unauthorized access and make the system vulnerable~\cite{ alevizopoulou2021social, gasmi2019information,vuldatPaper}. An \textbf{attack} is a way to exploit a vulnerability of a system, while several \textbf{Tactics} can be combined to represent an attack pattern for a system~\cite{theisen2018attack}.  
 Since it is impossible to predict whether a weakness exists when a vulnerability will be exploited or what impact an attack will have on a system, it is of foremost importance to provide stakeholders with instruments to detect and, possibly, remove vulnerabilities~\cite{iorga2021yggdrasil, queiroz2019eavesdropping,baccar2021automated, dionisio2019cyberthreat,tang2023csgvd}.
 
When a \textbf{vulnerability} is revealed and given a CVE identifier, the vendor organization in charge of the software starts working on a patch to fix the issue. The purpose of patching, along with other strategies for addressing coding flaws, is to identify and resolve problems before attackers can exploit them~\cite{dempsey2017automation}.
The CVE repository contains vulnerability reports from different systems and applications that can be publicly accessed~\cite{othman2024vulnerability, WhatCVE}. 
 For every publicly known vulnerability, a CVE report includes various pieces of information, including a description, an identification number (CVE-ID), and at least one reference to a system in which the vulnerability has been exploited~\cite{sun2023automatic}, together with known vulnerability fixes and mitigation, impact ratings, and severity scores based on the Common Vulnerability Scoring System (CVSS) ~\cite{CVSS}. Table~\ref{tab:cveDataset} shows an example of an attack and description for CWE, and CVE records.
 We can see the common vulnerability and exposure CVE-2022-4826 affecting the Simple Tooltips WordPress plugin prior to version 2.1.4. This vulnerability is related to improper validation and escaping of shortcode attributes before rendering them on a page or post. As a result, users with contributor roles and above could inject malicious scripts, leading to Stored Cross-Site Scripting (XSS) attacks. This vulnerability can be linked to the Pattern of \textbf{attack} CAPEC-38 with an attacker attempting to load a malicious resource into the program so that it will be executed.
 
 CWE is a community-developed collection of typical weaknesses in software, coding errors, and security flaws. A CWE report includes various information such as a name, an identification number (CWE-ID), a description, an observed example containing CVE reports ID related to CWE, and related attack patterns highlighting the relationships between specific attack patterns and CWE. CWE and CVE reports are connected through the CVE-ID. 

\subsection{Attack knowledge bases}
\label{sec:TTP}
\begin{figure*}[htb!]
\centering
% \begin{subfigure}[b]{0.4\textwidth}
% \includegraphics[scale=0.6]{Fig/Relationship MITRE ATTACK.drawio.pdf}
% \caption{ATT\&CK model.\label{fig:attackModel}}
% \end{subfigure}\hspace{8pt}
% \begin{subfigure}[b]{0.6\textwidth}
\includegraphics[scale=0.9]{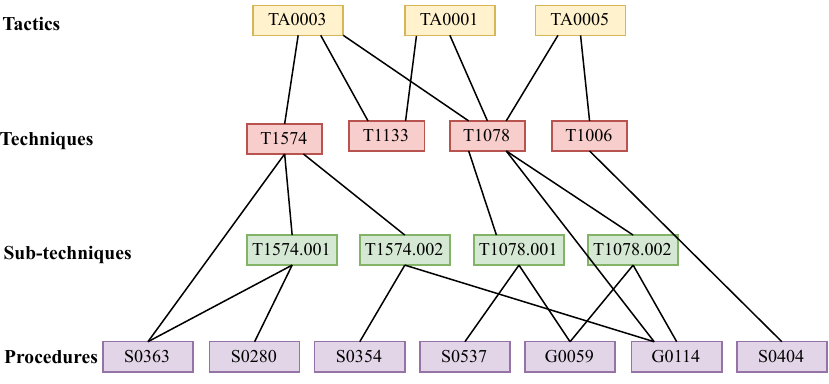}
\caption{An example of the ATT\&CK relations.}
\label{fig:mitrerelationship}
% \end{subfigure}
% \caption{The attack model.}\label{sec:attackmodel}
\end{figure*}

We leveraged the information on attacks contained in the ATT\&CK and CAPEC repositories.
ATT\&CK~\cite{ATTACK} is a framework that provides a comprehensive inventory of \textbf{tactics}, \textbf{techniques}, and \textbf{procedures} adopted by attackers during various stages of a cyberattack~\cite{rahman2024attackers,son2023introduction, irshad2023cyber}. TTPs help security experts understand adversary behavior, guide threat detection and response efforts, and enhance organizational defenses against cyberthreats. According to the ATT\&CK model, \textbf{tactics} represent the goal of an adversary's attack, like ways to get access to a secured network, whereas \textbf{techniques} define how to carry out an attack in the context of a certain Tactic, like using phishing attempts~\cite{MITRE, satvat2021extractor}. Techniques can be further detailed into \textbf{subtechniques}. \textbf{Procedures} provide detailed insights into how adversaries effectively use the techniques in the actual attacks. For example, a procedure could involve sending emails camouflaged as legitimate communications from a trusted source to trick users into clicking malicious links.
% A procedure is the simplest form of an attack's step-by-step execution.
Researchers typically use TTPs to profile or examine the attack life cycle on a specific system. 
Our study used Version 14 of the ATT\&CK framework, which contains 14 Tactics, 201 Techniques, 424 Subtechniques, and 809 procedures. The framework also provides the ATT\&CK Matrix, which breaks each attack down into tactics, techniques, and procedures. 
An example of the relations among Tactics, Techniques, Subtechniques, and Procedures are shown in Figure~\ref{fig:mitrerelationship}. All relations are many-to-many. For instance,  the Technique \textit{``event triggered execution"} has 16 Subtechniques, whereas, \textit{``account access removal, acquire access, and audio capture"} has no Subtechnique. 
The Common Attack Pattern Enumeration and Classification (CAPEC)~\cite{CAPEC} is a catalog of common \textbf{Attack Patterns}. CAPEC is linked to CWE using the CWE-ID, creating an association between specific attack techniques and the underlying weaknesses they exploit ~\cite{ATTACK}. Attack Patterns allow us to see what techniques in the ATT\&CK repository an adversary could use to target the weaknesses and vulnerabilities.
For instance, the Pattern\textit{ ``privilege escalation (CAPEC-233)"} elevates attackers' privilege to perform an action they are not supposed to be authorized to perform. Related to this Pattern, one can link the Technique \textit{``Abuse Elevation Control Mechanism (T1548)"} that describes the techniques that attackers use to gain higher-level permissions on a system or network. Among those, there is the Technique \textit{``Access token manipulation (T1134)"}, which can be implemented in Metasploit~\cite{Metasploit, Metasploitprocedure} with the procedure  \textit{``named-pipe impersonation"}~\cite{MITRE, wu2021price, noor2019machine}.
%The MITRE ATT\&CK is a repository that describes adversaries' tactics, techniques, and procedures during cyber attacks. 
We will connect ATT\&CK with CAPEC in our dataset, matching the techniques' IDs.

The attack descriptions in these repositories vary significantly in their structure, length, and level of detail. For example, ATT\&CK Tactics are abstract and short, while CAPEC Patterns are typically longer and more technically detailed. Table~\ref{tab:AttacsDescriptions} summarizes the typical characteristics of the four types of attack descriptions.
Manually linking 625 ATT\&CK techniques to over 295,000 CVEs is impractical. To address this, we employ sentence transformers to enable a fully automated and semantically driven integration across ATT\&CK, CAPEC, CWE, and CVE repositories.
\begin{table}[htb]
 \caption{Typical characteristics of the attack types.}
    \label{tab:AttacsDescriptions}
    \centering
    %\fontsize{9pt}{13pt}\selectfont
    \begin{tabular}{@{\extracolsep{\fill}}p{1.5cm}p{3.3cm}p{3.3cm}p{3.3cm}p{3.3cm}}
    \hline
&\textbf{Tactic}&\textbf{Technique}&\textbf{Procedure}         & \textbf{Pattern} \\\hline
\textbf{Definition}&               	 High-level goal of an adversary  &	 General method of how the adversary achieves a goal 	 &Specific instance of an attacker using a technique 	 &Structured description of a recurrent method of attack \\          \textbf{Vocabulary} &      	 Abstract, it uses mostly generic terms    &                  	 It is more specific, but it still uses reusable terms        	& It includes tool names, scripts, commands         &  	It uses  technical terms, exploit-specific language      \\                
 % \textbf{Length}&         	1–5 words                   &                       	  Sentence  &                            	 Multiple sentences              &          	 One or more paragraphs with structured data  \\
 \hline
    \end{tabular}
   
\end{table}

\begin{table*}[htb]
\caption{Overview of SOTA Sentence Transformer Models per architecture and ordered by embedding dimensions.}
\label{tab:pretrainedmodels}
\centering
 \fontsize{8pt}{13pt}\selectfont
\begin{tabular}{llccc}
\hline  
\textbf{Acronym}&\textbf{Model} & \textbf{Architecture} & \textbf{Embedding Dimension}  & \textbf{Model Size} \\
\hline
PAlbert	&	paraphrase-albert-small-v2\footnotemark[1] 	&	 ALBERT 	&	768	&	 43 \, MB \\
PTinyBERT	&	paraphrase-TinyBERT-L6-v2\footnotemark[2] 	&	 TinyBERT 	&	768	& 240  MB \\
MDBERT	&	multi-qa-distilbert-cos-v1\footnotemark[3] 	&	 DistilBERT 	&	768	&	 250 MB \\
MSMBERT	&	msmarco-bert-base-dot-v5\footnotemark[4] 	&	 BERT 	&	768	&	 420 MB \\
DRoBERTa	&	all-distilroberta-v1\footnotemark[5] 	&	 DistilRoBERTa 	&	768	&	 290 MB \\
Roberta	&	all-roberta-large-v1\footnotemark[6] 	&	 RoBERTa 	&	1024	&	 1.36 GB \\
\hline
MiniLM6	&	all-MiniLM-L6-v2\footnotemark[7] 	&	 MiniLM 	&	384	&	 80 \, MB \\
MiniLM12	&	all-MiniLM-L12-v2\footnotemark[8] 	&	 MiniLM 	&	384	&	 120 MB \\
MMiniLM6	&	multi-qa-MiniLM-L6-cos-v1\footnotemark[9] 	&	 MiniLM 	&	384	&	 80 \, MB \\
PMiniLM6	&	paraphrase-MiniLM-L6-v2\footnotemark[10] 	&	 MiniLM 	&	384	&	 80 \, MB \\
PMiniLM12	&	paraphrase-multilingual-MiniLM-L12-v2\footnotemark[11] 	&	 MiniLM 	&	384	&	 420 MB \\
\hline
MPNet	&	all-mpnet-base-v2\footnotemark[12] 	&	 MPNet 	&	768	&	 420 MB \\
MMPNet	&	multi-qa-mpnet-base-dot-v1\footnotemark[13] 	&	 MPNet 	&	768	&	 420 MB \\
\hline
XXLT5	&	gtr-t5-xxl\footnotemark[14] 	&	 T5-XXL 	&	4096	&	 9.23 GB \\

\hline
\end{tabular}%
\end{table*}

\subsection{Sentence transformers}\label{sec:transformer}
Sentence transformer models are pre-trained models that produce sentence embeddings for various natural language processing tasks, such as semantic search, paraphrasing, and clustering. However, these models often contain millions of parameters, making fine-tuning and deployment challenging due to latency and capacity constraints. 
\textcolor{blue}{In this work, we utilize 14 SOTA pre-trained sentence transformer models, as summarized in Table~\ref{tab:pretrainedmodels}. These models are recognized as top performers in semantic similarity~\cite{muennighoff2022mteb,colangelo2025comparative, choi2021evaluation, pretrained}. All models are sourced from the Hugging Face Model Hub~\footnotemark[1], \footnotetext[1]{\url{Hugging Face: https://huggingface.co/}}which provides a comprehensive repository of transformer-based models for natural language processing tasks.
The selected models have been extensively benchmarked and fine-tuned for sentence-level tasks, such as semantic textual similarity, information retrieval, and clustering~\cite{muennighoff2022mteb,colangelo2025comparative, choi2021evaluation, pretrained}. Notably, models like MPNet, MiniLM, MSMARCO, RoBERTa, and T5 consistently achieve strong results across benchmarks~\cite{ muennighoff2022mteb, colangelo2025comparative}. Lightweight models, including PAlbert, have also demonstrated competitive performance in the evaluations~\cite{choi2021evaluation}. 
A summary table listing these models along with their architecture, embedding size, and benchmark performance for semantic similarity is available in the official sentence transformers documentation~\cite{pretrained}, supporting the transparency and reproducibility of our model selection.
In addition, we selected models with different architectures and examined how these architectural differences affect performance in linking attack descriptions to CVEs.}
% we utilize 14 SOTA pre-trained sentence transformer models~\cite{pretrained}, as summarized in Table~\ref{tab:pretrainedmodels}. These models are recognized among the top performers for semantic similarity, sentence embeddings, and semantic search. All the models are sourced from Hugging Face\footnotemark[15], which provides a comprehensive collection of high-quality transformer-based models for natural language processing tasks. Specifically, they have been extensively benchmarked and fine-tuned for tasks involving sentence-level representations, such as semantic textual similarity, information retrieval, and clustering.
% In this work, we utilize 14 SOTA of the pre-trained sentence transformer models~\cite{pretrained} as summarized in Table \ref{tab:pretrainedmodels}
% , which are among the top-performing models for semantic similarity, sentence embeddings, and semantic search. These models have been extensively benchmarked and optimized for tasks involving sentence-level representations, such as semantic textual similarity, information retrieval, and clustering, which is sourced from Hugging Face~\cite{huggingface}.
% These models are built on architectures, including BERT~\cite{devlin2018bert}, MPNet ~\cite{song2020mpnet}, DistilBERT ~\cite{sanh2019distilbert}, RoBERTa ~\cite{liu2019roberta}, MiniLM ~\cite{wang2020minilm}, and T5~\cite{raffel2020exploring}. 

BERT~\cite{reimers2019sentence} is a pre-trained transformer model designed for natural language understanding. It is based on a deep bidirectional architecture, capturing context from both left and right of a word, which significantly enhances performance on tasks like text classification, question answering, and more. 
The BERT-based models include  \texttt{MSMBERT}, optimized for semantic search, as well as \texttt{PAlbert} and \texttt{PTinyBERT}, which are smaller versions that utilize parameter reduction techniques while still maintaining competitive performance.

\hspace{-7em}
\footnotetext[1]{\url{PAlbert: https://huggingface.co/sentence-transformers/paraphrase-albert-small-v2}}
\footnotetext[2]{\url{PTinyBERT: https://huggingface.co/sentence-transformers/paraphrase-TinyBERT-L6-v2}}
\footnotetext[3]{\url{MDBERT: https://huggingface.co/sentence-transformers/multi-qa-distilbert-cos-v1}}
\footnotetext[4]{\url{MSMBERT: https://huggingface.co/sentence-transformers/msmarco-bert-base-dot-v5}}
\footnotetext[5]{\url{DRoBERTa: https://huggingface.co/sentence-transformers/all-distilroberta-v1}}
\footnotetext[6]{\url{Roberta: https://huggingface.co/sentence-transformers/all-roberta-large-v1}}
\footnotetext[7]{\url{MiniLM6: https://huggingface.co/sentence-transformers/all-MiniLM-L6-v2}}
\footnotetext[8]{\url{MiniLM12: https://huggingface.co/sentence-transformers/all-MiniLM-L12-v2}}
\footnotetext[9]{\url{MMiniLM6: https://huggingface.co/sentence-transformers/multi-qa-MiniLM-L6-cos-v1}}
\footnotetext[10]{\url{PMiniLM6: https://huggingface.co/sentence-transformers/paraphrase-MiniLM-L6-v2}}
\footnotetext[11]{\url{PMiniLM12: https://huggingface.co/sentence-transformers/paraphrase-multilingual-MiniLM-L12-v2}}
\footnotetext[12]{\url{MPNet: https://huggingface.co/sentence-transformers/all-mpnet-base-v2}}

\footnotetext[13]{\url{MMPNet: https://huggingface.co/sentence-transformers/multi-qa-mpnet-base-dot-v1}}
\footnotetext[14]{\url{XXLT5: https://huggingface.co/sentence-transformers/gtr-t5-xxl}}

DistilBERT and DistilRoBERTa, on which \texttt{MDBERT} and \texttt{DRoBERTa} are based, compress BERT and RoBERTa, respectively, using knowledge distillation, reducing model size by 40\% while retaining 97\% of their accuracy, making them suitable for real-time applications. RoBERTa enhances BERT by removing next-sentence prediction and training on more data, leading to better generalization.
Additionally, MiniLM architecture~\cite{wang2020minilm} models such as \texttt{MMiniLM6}, \texttt{MiniLM6}, 
\texttt{PMiniLM12}, \texttt{MiniLM12}, and \texttt{PMiniLM6}, 
% ~\cite{paraphrase-MiniLM-L6-v2}, 
further reduce computational complexity by employing a minimalistic self-attention mechanism. These models are a new version of the BERT model ~\cite{reimers2019sentence}, a popular model for natural language understanding, that has been created using knowledge distillation~\cite{HintonEtAl2015} to compress the large model BERT (called the teacher model) into a small model MiniLM (called the student model), which uses much fewer parameters and computations while achieving competitive results on downstream tasks. MiniLM notably reduces the effort for finding the most similar pairs while maintaining the accuracy of BERT. Moreover, they employ a 6 and a 12-layer neural network to generate vectors that encapsulate the text. Unlike BERT, which learns from a single language, these models are trained on a diverse dataset from 50 different languages.
The MPNet-based model, \texttt{MMPNet} and \texttt{MPNet}, extend BERT by integrating XLNet’s permutation-based training, improving contextual understanding and retrieval accuracy.
T5~\cite{raffel2020exploring}, exemplified by \texttt{XXLT5} model, is designed with a transformer architecture and operates on large-scale text-to-text tasks, allowing it to be used for summarization, translation, classification, and more, making it highly versatile across applications.
In this study, we use pre-trained transformer models to generate embeddings for attack and vulnerability descriptions, and then apply a similarity layer based on cosine similarity to determine whether an attack is linked to a vulnerability (Section~\ref{sec:TuningEmbedding}).
\begin{figure}[t!]
\centerline{\includegraphics[width=\linewidth]{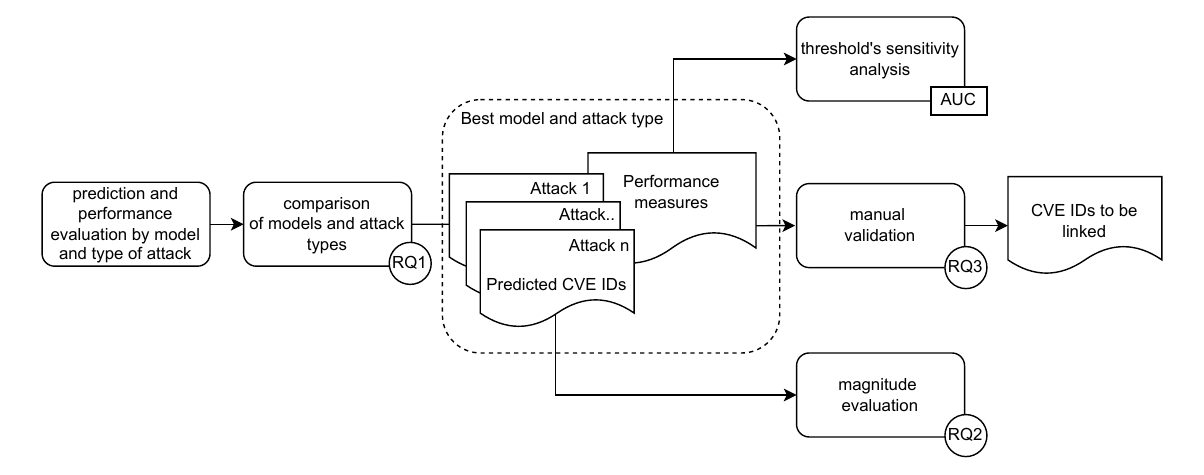}}
\caption{\textcolor{blue}{Overview of the methodology for linking attack descriptions to CVE reports.}}
\label{fig:pipeline}
\end{figure}
\section{Methodology}
\label{sec:comparisonStufy}

Figure~\ref{fig:pipeline} overviews our approach. Firstly, for each model  (sentence transformer) and type of attack text (Tactic, Technique, Procedure, or Attack Pattern), we perform a \textit{vulnerability prediction and model's performance analysis} as described in Section~\ref{sec:Prediction}, \textcolor{blue}{to generate ranked lists of potential CVE matches and quantify model effectiveness}. 
Secondly, to answer RQ$_1$, we compare models and attack types by their performance on the corresponding attacks as illustrated in Section~\ref{sec:Comparison}, \textcolor{blue}{identifying the model–attack-type combination that achieves the highest prediction accuracy}. Then, with the best model and attack type, we perform 1) a \textit{sensitivity analysis} of the threshold used to determine the predicted vulnerabilities as illustrated in Section~\ref{sec:SensitivityAnalysis}, \textcolor{blue}{to identify the threshold value that yields the best trade-off between precision and recall}, 2) an evaluation of the number of correctly predicted vulnerabilities per attack text (\textit{magnitude evaluation}) to answer RQ$_2$ in Section~\ref{sec:Magnitude}, and 3) a \textit{manual validation} of the predicted vulnerabilities to discuss RQ$_3$, Section~\ref{sec:Validation}, \textcolor{blue}{ to demonstrate their practical value for expanding MITRE’s ATT\&CK-CVE mappings and reducing manual effort}.

\subsection{Vulnerability prediction and model's performance analysis}\label{sec:Prediction}
\begin{figure}[t!]
    \centering
    \includegraphics[width=\linewidth]{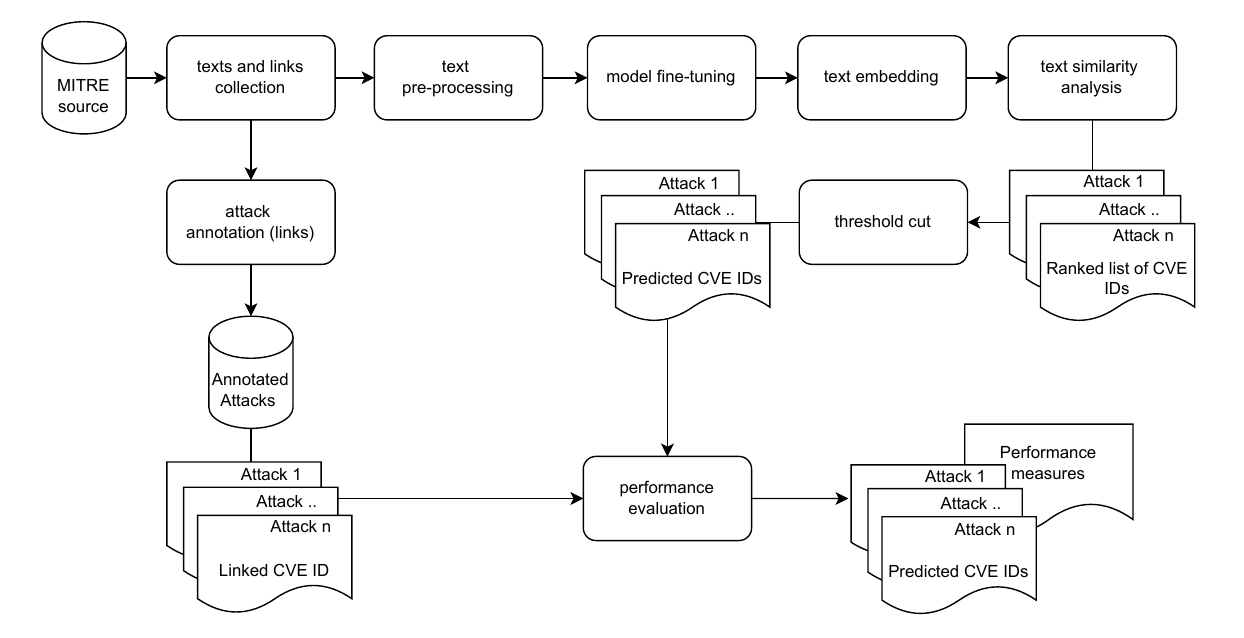}
    \caption{
    \textcolor{blue}{Workflow for prediction and performance evaluation by model and attack type.}}
    \label{fig:Prediction}
\end{figure}

Figure~\ref{fig:Prediction} shows how we use a sentence transformer to predict vulnerabilities from an attack text of a given type. We collect all attack descriptions related to that type, along with all vulnerability descriptions and their IDs (\textit{texts and links collection}). 
For each attack ID, we associate the CVE IDs eventually found in the attack's description page (links) (\textit{attacks annotation}). 
Then, we pre-process all texts  (\textit{text pre-processing}). After fine-tuning the model, we embed all texts with the selected model  (\textit{text embedding}). 
The embedding of each attack text is compared for similarity with the embeddings of all vulnerabilities (\textit{text similarity analysis}).
Vulnerability embeddings that have a similarity score greater than the threshold with a given attack embedding are
labeled as predicted vulnerabilities for that attack.
In this way, we obtain a mapping between the attack IDs and  CVE IDs. 
At this point, we evaluate the performance of the sentence transformer by comparing the sets of predicted and linked CVE IDs. 
In the next step, we compare the performance by varying the model and attack type. In the following, we describe this process in more detail.
\subsubsection{Texts and
links collection}
\label{sec:DataCollection}
\textcolor{blue}{This step systematically collects and structures attack descriptions and vulnerability records from MITRE repositories to provide a transparent, reproducible foundation and ground truth for evaluating the automated linking approach.}
From the MITRE repositories, we first collect the descriptions of the attacks of all types as well as the descriptions of all vulnerabilities. By inspecting the repositories' web pages, we also retrieve any links that explicitly connect attacks and vulnerabilities (\textit{explicit link}).
%Example CWE787 Note: this is a curated list of examples for users to understand the variety of ways in which this weakness can be introduced. It is not a complete list of all CVEs that are related to this CWE entry.
For example, on the web page of the weakness CWE-770 there exist 10 different explicit links to CVEs and we map CWE-770 to the corresponding CVE IDs.  
\begin{figure*}[hbt!]
% \centerline{\includegraphics[width=0.75\columnwidth]{Fig/MITRESources3.drawio.pdf}}
\centerline{\includegraphics[width=0.75\columnwidth]{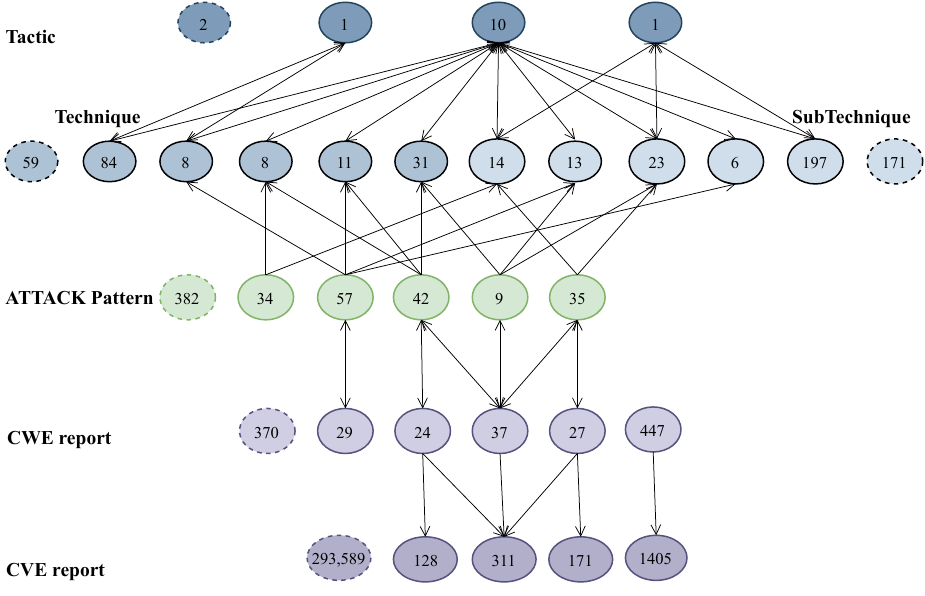}}
\caption{
\textcolor{blue}{
Graph representation of explicit links between attack types, weaknesses, and vulnerabilities in MITRE repositories, showing the density and direction of existing connections.}
% Explicit links between data in MITRE repositories.
}
\label{fig:graphM}
\end{figure*} 
The graph in Figure~\ref{fig:graphM} illustrates the result of such analysis.  Each node clusters attacks, weaknesses and vulnerabilities, and its value indicates the number of explicit links found in the respective web pages of a node of another type. 
\textcolor{blue}{
Some nodes in the graph remain isolated, referred to as “floating entries”, and the majority of CVE reports are represented within this category. 
The portion of floating entries somehow measures the lack of knowledge about the relation between attacks and vulnerabilities. 
On the other hand, there are “Super Entries”, nodes with many outgoing or incoming edges. The portion of these nodes represents the more mature knowledge of attacks, vulnerabilities, and their relations.}
In addition, the directions of the graph edges indicate where the links have been found. 
For instance, there are 447 CWE web pages that include links to 1405 CVEs. As Procedures are only listed in the Techniques' pages,  there are also no explicit links between Patterns or Tactics and Procedures. Therefore, we omit the Procedures' nodes.
Table~\ref{tab:golddataset} reports the number of CVEs and CWEs that are linked or not to attacks.
\begin{table*}[htb!]
% \scriptsize
\caption{Number of CVEs and CWEs linked and not linked to attacks.}
\label{tab:golddataset}
\centering
\bgroup
\begin{tabular}{llccccc}
\hline
 & &\textbf{Tactic} &\textbf{Technique} &\textbf{Procedure} & \textbf{Attack Pattern} \\
\hline
\textbf{CWE} & linked& 117&79& 117& 117  \\
 & not linked &818&856&818&818\\
\hline
\textbf{CVEs} &linked&610&610& 610& 610 \\
&not linked &294994&295165&294994 &294994 \\
\hline
\end{tabular}%
\egroup
\end{table*}
In particular, the table shows that a large majority of vulnerabilities are not explicitly linked in any attack page. This may be due to the fact that we leverage the explicit links between CWE and CVE to link attacks to vulnerabilities. As noted in the CWE pages, the explicit links to CVEs is  \textit{``a curated list of examples for users to understand the variety of ways in which this weakness can be introduced. It is not a complete list of all CVEs that are related to this CWE entry."} Thus, some existing links may not be reported in the page. 

% In this section, we introduce the mapping $\mathcal{M}$ between the attacks and CVE reports that can be retrieved from the ATT\&CK and CAPEC and CVE repositories, respectively. The mapping will serve as ground truth to assess the performance of our approach.
% $$\mathcal{M}: A\rightarrow C \\$$
% $$\mathcal{M}(a) =\{ c \in \mathcal{C}   \, : \, \exists  \, a \rightarrow c\}$$
% where $C$ is the set of all CVE reports, $A$ is the set of all attack reports in ATT\&CK and CAPEC.
% The mapping associates an attack text, $a$, to a list of  CVE reports through a \textit{link} ($\rightarrow$) derived from the information contained in the repositories ATT\&CK, CAPEC, CWE, and CVE. 
%To determine the link in the mapping, we first reviewed the \textit{explicit links}  existing in the reports' web pages of such repositories. 
%\enlargethispage{\baselineskip}
%
\subsubsection{Attacks' annotation}\label{sec:AttacksAnnotation}
We utilize explicit links to establish connections between attacks and a defined set of vulnerabilities.
As illustrated in Figure~\ref{fig:graphM}, there are no explicit links between Tactics, Techniques, or Procedures and CWEs or CVEs.  
We then exploit the explicit links between CWEs and attack patterns to annotate attacks with the CVEs to which they are linked. 
\textcolor{blue}{This step creates a traceable mapping between attacks and vulnerabilities, forming the ground truth for model evaluation.}
For instance, given a Technique, we retain in CAPEC all attack patterns that mention it. For each Attack Pattern, we then collect all the CWEs whose explicit link mention it, and from those CWEs, we retrieve the explicit links to CVEs. 
Following this procedure, Table~\ref{tab:golddataset2} illustrates the attacks that are linked to some CVEs. 
\begin{table*}[htb]
% \scriptsize
\caption{Attacks linked and not linked to CVE reports.}
\label{tab:golddataset2}
\centering
\bgroup
\def\arraystretch{1}% 1 is the default, change whatever you need
\setlength{\tabcolsep}{4pt}
\begin{tabular}{lccccccc}
\hline  
& \textbf{Tactic} &\textbf{Technique} &\textbf{Procedure} & \textbf{Attack Pattern} \\
\hline
 \textbf{Linked}&11 & 100  & 721 &86\\
\textbf{Not linked}&3&525&88&473\\
\hline
\textbf{Total}&14&625&809&559\\
\hline
\end{tabular}%
\egroup
\end{table*} 
Table~\ref{tab:cveDataset} illustrates an example of such links in the case of the Technique T1574.007, Attack Pattern CAPEC-38, CWE-427, and CVE-2022-4826. The table includes only one representative instance among all those that can be linked to Technique T1574.007 using our procedure.
 \begin{table*}[htbp!]%
\centering
% \small
\caption{An example of a Technique and its chain to one of the linked CVEs.}%
\label{tab:cveDataset}
\begin{small}
\begin{tabular*}{\textwidth}{@{\extracolsep{\fill}}p{3.5cm}p{3.5cm}p{3.5cm}p{3.5cm}}%
\hline
\textbf{Technique-T1574.007:} Adversaries may execute their own malicious payloads by hijacking environment variables used to load libraries. Adversaries may place a program in an earlier entry in the list of directories stored in the PATH environment variable, which Windows will then execute when it searches sequentially through that PATH listing in search of the binary that was called from a script or the command line.

% The PATH environment variable contains a list of directories. Certain methods of executing a program (namely using cmd.exe or the command-line) rely solely on the PATH environment variable to determine the locations that are searched for a program when the path for the program is not given. If any directories are listed in the PATH environment variable before the Windows directory, <code>%SystemRoot%\system32</code> (e.g., <code>C:\Windows\system32</code>), a program may be placed in the preceding directory that is named the same as a Windows program (such as cmd, PowerShell, or Python), which will be executed when that command is executed from a script or command-line.

% For example, if <code>C:\example path</code> precedes </code>C:\Windows\system32</code> is in the PATH environment variable, a program that is named net.exe and placed in <code>C:\example path</code> will be called instead of the Windows system "net" when "net" is executed from the command-line.
& \textbf{CAPEC-38:} This Pattern of attack sees an adversary load a malicious resource into a program's standard path so that when a known command is executed then the system instead executes the malicious component. The adversary can either modify the search path a program uses, like a PATH variable or classpath, or they can manipulate resources on the path to point to their malicious components. 
% J2EE applications and other component based applications that are built from multiple binaries can have a very long list of dependencies to execute. 
If one of these libraries and/or references is controllable by the attacker then application controls can be circumvented by the attacker.
&\textbf{CWE-427:} The product searches for critical resources using an externally-supplied search path that can point to resources that are not under the product's direct control.
& \textbf{CVE-2022-4826:} The Simple Tooltips WordPress plugin before 2.1.4 does not validate and escape some of its shortcode attributes before outputting them back in a page/post where the shortcode is embed, which could allow users with the contributor role and above to perform Stored Cross-Site Scripting attacks.
\\\hline
\end{tabular*}
\end{small}
\end{table*}
Finally, we annotate each attack ID by the CVE IDs we link with our procedure. For instance, in the above example, Technique ID T1574.007 is annotated with CVE-2022-4826. In the general case, an attack ID can be annotated with several CVEs IDs. The annotated dataset can be found in our replication package~\cite{VULDATDataSet} and is used as ground truth for the performance analysis in Section~\ref{sec:PerformanceEvaluation}. 
Using this annotation, for each attack ID $a,$ we build the \textit{set of linked CVE IDs}: 
 $$\mathcal{M}(a) =\{ c \in \mathcal{C}   \, : \, \exists  \, a \rightarrow c\}$$
 where $C$ is the set of all CVE IDs and $a \rightarrow c$ indicates a link between the attack ID $a$ and the CVE ID $c$ found with our procedure.
%
% BR I have removed this table as it's not necessary, it's already included in the large figure
%  \begin{figure}[htp!]
% \centerline{\includegraphics[width=0.75\columnwidth]{Fig/textPreprocessing.drawio2.pdf}}
% % \centerline{\includegraphics{Fig/textPreprocessing.drawio2.pdf}}
% \caption{Text pre-processing.}
% \label{fig:textpreprocessing}
% \end{figure}
% \subsection{Text pre-processing}
% \label{sec:textpreprocessing}
% We pre-processed the attack and CVE descriptions to remove irrelevant information and noise as illustrated in Figure.~\ref{fig:textpreprocessing}. 
% We started by lowering cases and removing citations and URLs. Then, we performed in our experiment the same pre-processing steps as those automatically executed by the pretrained model: tokenization, stemming, lemmatization, stop word removal, and handling punctuation.
% Tokenization, stemming, and punctuation removal are critical steps. Tokenization splits the text into tokens or single words, allowing us to handle the text more effectively and extract meaningful insights and patterns from the data more easily.
% Stemming is the process of breaking down a word into its stem or root form. This process aims to unify variations of words that share the same core meaning, reduce the complexity of the data, and improve our analysis.
\subsubsection{Text pre-processing}\label{sec:Preprocessing}
\textcolor{blue}{
We pre-processed the attack and CVE descriptions to remove irrelevant information and textual noise while retaining semantic content. The text was normalized to lowercase, and citations, URLs, and general non-alphanumeric characters were removed. Unlike traditional NLP pipelines, we did not apply stemming, lemmatization, or stop-word removal, as prior studies~\cite{okonkwo2023leveraging, siino2024text} have shown that transformer-based models rely on subword tokenization and contextual embeddings, which benefit from preserving grammatical structure and functional words. Stop words, in particular, contribute to the syntactic flow of sentences and often enhance the contextual understanding that attention-based models exploit. Similarly, avoiding stemming and lemmatization ensures that morphological variants are preserved, allowing the model to capture subtle differences in meaning. This minimal preprocessing balances noise reduction with semantic preservation, enabling the sentence transformer to better capture relationships between attack descriptions and CVE reports.
}
\subsubsection{Model fine-tuning and text embedding }\label{sec:TuningEmbedding}
We have implemented our approach using 14 pre-trained sentence transformer models (as discussed in Section~\ref{sec:transformer}),
\textcolor{blue}{with embedding sizes ranging from 384 to 4096 dimensions. These models are designed to produce sentence embeddings}, fixed-size vector representations of input texts.
Specifically, both attack and vulnerability descriptions are transformed into vectors in a shared vector space, allowing for their comparison. 
Each of the selected models uses a multi-layer architecture to process input text. The tokens of each text are passed through multiple transformer encoder layers, which employ self-attention mechanisms and feed-forward networks to capture the contextual relationships among the tokens.  Afterward, a pooling strategy such as mean pooling or using the [CLS] token is applied to condense the token-level embeddings into a fixed-length sentence embedding. 
\textcolor{blue}{For model fine-tuning, we applied a single split to our dataset into three subsets with ratios 80\%, 10\%, and 10\% and apply the model for training, validation, and testing, respectively, as recommended in~\cite{HinidumaEtAl2025}. This fixed split was chosen to ensure fair comparability between models, avoid variability from multiple random splits, and maintain reproducibility. Dataset statistics before splitting, including linked and unlinked samples for each attack type, are reported in Table~\ref{tab:golddataset2}. 
All models were fine-tuned using CosineSimilarityLoss. Training was conducted for four epochs with 100 warmup steps, and evaluation was performed every 500 steps, following established practices in sentence transformer training examples~\cite{hyperparams}.
The validation set was used for intermediate evaluation, and the test set for final performance reporting.}
% \subsubsection{Computational Environment}
% \label{sec:compEnv}
% \textcolor{blue}{
% We conducted the experiments on a server equipped with a 64-core Intel® Xeon® Gold 6246R CPU @ 3.40 GHz, 92 GiB of main memory, and high-capacity storage. The system ran Ubuntu 22.04.4 LTS (Jammy).}

% As an example, Figure~\ref{fig:BERT} shows the architecture of the BERT model~\cite{mamede2022exploring}.
% \begin{figure}[hbt!]
%     \centering
%     \includegraphics[width=0.3\textwidth, trim={0cm 1.5cm 0cm 1.5cm}, clip]{Fig/BERTARCH.drawio.pdf}
%     \caption{BERT architecture.}
%     \label{fig:BERT}
%     \index{figures}
% \end{figure}
% \begin{figure}[hbt!]
%         \centering
%         \includegraphics[width=.4\textwidth]{Fig/BERTARCH.drawio.pdf}
%         \caption{BERT architecture \cite{mamede2022exploring}}
%         \label{fig:BERT}
%         \index{figures}
%         \end{figure}
\subsubsection{Similarity analysis}\label{sec:Similarity}
\textcolor{blue}{Following established practices in embedding-based NLP systems~\cite{reimers2019sentence,muennighoff2022mteb}, we adopt cosine similarity as the primary metric for comparing sentence embeddings in transformer-based models. This choice aligns with its well-established role in embedding-based machine learning systems, where it has consistently demonstrated effectiveness across a range of natural language processing tasks. For instance, Reimers and Gurevych~\cite{reimers2019sentence} showed that applying cosine similarity in Sentence-BERT significantly improved performance in semantic textual similarity benchmarks, while Muennighoff et~al.~\cite{muennighoff2022mteb} adopted cosine similarity as the standard metric across more than 50 models, validating its robustness for tasks such as information retrieval, clustering, and semantic search.} In this work, we perform cosine similarity (Equation (\ref{eqSim})) on the normalized output vectors of an attack embedding $p$ = ($p1$, $p2$, ..., $pn$) with all CVE embeddings $q$ = ($q1$, $q2$, ..., $qn$). 
Cosine similarity is a common metric in natural language processing and information retrieval because it captures the angular relationship between vectors, focusing on their direction rather than magnitude. This property makes it especially effective for comparing sentence embeddings, where semantic meaning is primarily represented by the vector's orientation in the embedding space.
\textcolor{blue}{Cosine similarity scores theoretically range from -1 to 1, where 1 indicates maximum similarity, 0 indicates no correlation, and -1 indicates complete opposition. Since we normalize embeddings and focus exclusively on semantically related pairs, our practical range of interest is [0, 1].
For consistency with the threshold selection procedure, these values are expressed on a 0–100 scale in our experiments.
This property makes cosine similarity especially effective for comparing sentence embeddings, where semantic meaning is primarily represented by the vector’s orientation in the embedding space.}
% The similarity score ranges in $[0,1]$ with 1 maximum similarity and 0 no similarity.
\begin{equation}
\label{eqSim}
Sim(\vec{p} \cdot \vec{q}) = \frac{\vec{p} \cdot \vec{q}}{\|\vec{p}\| \cdot \|\vec{q}\|} = \frac{\sum_{i=1}^{n} p_i q_i}{\sqrt{\sum_{i=1}^{n} p_i^2} \cdot \sqrt{\sum_{i=1}^{n} q_i^2}}
\end{equation}
We rank all CVEs' embeddings by the similarity score with the attack embedding. 
A threshold $\rho$ is applied to cut the resulting ranked list at a certain level of similarity. 
Thus, we associate to an attack ID $a$ a list of CVE IDs whose  similarity score is  greater than the threshold:
$$\mathcal{L}_{\rho}(a) = \{ c \in C \,:\, Sim(\vec{a} \cdot \vec{c}) > \rho\}
$$ where $C$ is the set of all CVE IDs.  $\mathcal{L}_{\rho}(a)$ is the set of \textit{predicted CVEs  IDs} for an attack ID $a$ and a threshold $\rho$. 
 With $\mathcal{L}_{\rho}$, we can define a classification problem: a model predicts a vulnerability from an attack if the set $\mathcal{L}_{\rho}(a)$ for the attack ID $a$ is non-empty and it does not if the set is empty.
\subsubsection{Performance evaluation}
\label{sec:PerformanceEvaluation}
We can evaluate the performance of a model on the above classification problem. A perfect model should satisfy the following condition for any attack ID $a$: either the intersection between the predicted and the linked CVE IDs is non-empty,
$$\mathcal{L}_{\rho}(a) \cap \mathcal{M}(a) \neq \emptyset,
$$
or both sets are empty:
$$\mathcal{L}_{\rho}(a) = \emptyset \, \land \, \mathcal{M}(a) = \emptyset.
$$  
Table~\ref{tab:postivesNegatives} defines the classification problem in the attack set. 
% We denote the positives and negatives as follows.
\begin{table}[htbp!]
\caption{Classification in the attacks' set $\mathcal{A}$.}
\label{tab:postivesNegatives}
\begin{center}
% Adjusting table width to 0.5\textwidth
\begin{tabular}{ll}
\hline
% \begin{table}[htbp!]%
% % \scriptsize
% \caption{Positives and Negatives.}%
% \label{tab:postivesNegatives}
% \begin{center}
% \begin{tabular*}{0.5\textwidth}{@{\extracolsep{\fill}}ll}
\hline
\textbf{Type}  & \textbf{Description}  \\
\hline
% Positives & $\{ a \in \mathcal{A} \, : \exists \, {c} \in (\mathcal{C} \cap  \mathcal{M}_{a} )\}$\\
Positives & $\{ a \in \mathcal{A} \, : \exists \, {c} \in \mathcal{M}(a) \}$\\
Negatives &  $\{ a \in \mathcal{A} \, : \nexists \, {c} \in \mathcal{M}(a) \}$\\
Predicted Positives & $\{ a \in \mathcal{A} \, : \exists \, {c} \in \mathcal{L}_{\rho}(a) \}$\\
Predicted Negatives &  $\{ a \in \mathcal{A} \, : \nexists \, {c} \in \mathcal{L}_{\rho}(a) \}$\\
% True Positive ($a$) & $\left| \mathcal{L}_{a} \cap \mathcal{M}(a) \right| > 0$ \\
True Positives  (TP)& $\{ a \in \mathcal{A} \, :  \mathcal{L}_{\rho}(a) \cap \mathcal{M}(a) \not = \emptyset   \} $\\
% False Positive ($a$) &  $\left| \mathcal{L}_{a} - \mathcal{M}(a) \right| > 0$\\
False Positives  (FP) & $\{ a \in \mathcal{A} \, :\left( \mathcal{L}_{\rho}(a) \not = \emptyset   \, \land \, \mathcal{M}(a)  = \emptyset  \right)  \}$ \\
False Negatives (FN)  & $\{ a \in \mathcal{A} \, : \left( \mathcal{L}_{\rho}(a)  = \emptyset    \, \land \,  \mathcal{M}(a)  \not = \emptyset   \right)\}$ \\
True Negatives (TN)  & $\{ a \in \mathcal{A} \, : \left( \mathcal{L}_{\rho}(a)   = \emptyset   \,\land \,  \mathcal{M}(a)  = \emptyset   \right)\}$ \\

\hline
\end{tabular}
\end{center}
\end{table}
We compute the performance metrics by means of the cardinality of the sets in Table~\ref{tab:postivesNegatives}.
Being our approach a classifier, we use Precision, Recall, and their harmonic mean F$_1$ to evaluate its performance: 
%Precision is a metric for how well a classifier performs in recognizing instances of positive.  
\begin{equation}
\textit{Precision} = \frac{\textit{TP}}{\textit{TP} + \textit{FP}}
\end{equation}
%Precision measures the portion of correctly predicted CVE reports that are relevant.
\begin{equation}
% \begin{align*}
\textit{Recall} = \frac{\textit{TP}}{\textit{TP} + \textit{FN}}
% \end{align*}
\end{equation}
\begin{equation}
\textit{F$_1$} = 2 \times \frac{\textit{Precision} \times \textit{Recall}}{\textit{Precision} + \textit{Recall}}
% \end{align*}
\end{equation}
The F$_1$-score indicates how well the categorization task performs in terms of both precision and recall: the higher the F$_1$-score, the better. 
We fine-tune each of the pre-trained models on our dataset. To this aim, \textcolor{blue}{the same 80/10/10 train, validation and test split described in Section~\ref{sec:TuningEmbedding} was applied to all models to ensure consistency and a fair comparison.} 
For each attack type, we evaluate the performance of each model in the testing set using Precision, Recall, and F$_1$-Score. 
\textcolor{blue}{The experiments are conducted on a six-node GPU cluster, where each node is equipped with an NVIDIA A100 (80 GB), 192 GB of RAM, and an Intel Xeon 4208 (16-core) CPU. The code was implemented using PyTorch 2.8.0 and Hugging Face Transformers 4.55.2.}
\subsection{Comparison of models and attack types}\label{sec:Comparison}
To answer RQ$_1$, we compute Precision, Recall and F$_1$ as described in Section~\ref{sec:PerformanceEvaluation}, for each of the 14 models in Table~\ref{tab:pretrainedmodels} and all attack types (Tactic, Technique, Procedure and Pattern). 
\textcolor{blue}{In addition, this evaluation is to test our hypothesis that Technique descriptions are more informative than other attack types for vulnerability detection. Moreover, we hypothesize that transformer models with larger capacity and richer semantic representations will outperform lighter architectures.}
Finally, we use the maximum value of F$_1$ to determine the best model(s) and the attack type(s) on which the model(s) predict best. The output is pairs (attack type, model) where the model best performs on the set of that type of attack.

\subsection{Threshold's sensitivity analysis}
\label{sec:SensitivityAnalysis}
To identify the predicted CVE IDs (Section~\ref{sec:Prediction}), we initially set the similarity threshold $\rho = 60$\% based on our initial manual understanding of the samples.  
To validate our choice, we therefore perform a \textit{sensitivity analysis} by varying the threshold $\rho$ and re-evaluating the model's performance as recommended in ~\cite{lobo2022cost,sheng2006thresholding}. We select the best (attack type, model) pair identified in the comparison analysis in Section~\ref{sec:Comparison}. Then, we randomly sample a balanced subset from the original dataset of that attack type, consisting of 50 positive and 50 negative examples, on which we study the classification problem.
We employ the Receiver Operating Characteristic (ROC) curve to plot the True Positive Rate (TPR)
% $$TPR=\frac{TP}{TP+FN}$$
\begin{equation}
TPR=\frac{TP}{TP+FN}
\end{equation}
vs. the False Positive Rate (FPR)
\begin{equation}
FPR =\frac{FP}{TN+FP}
\end{equation}
across a range of threshold values from 1 to 100.  The threshold for which the ROC value achieves the closest value (Euclidean distance) to the values (0,1) - perfect classification -  will be selected.
The Area Under the ROC Curve (AUC) serves as a comprehensive measure of model performance across all thresholds, where a value of 1.0 indicates perfect classification, and 0.5 corresponds to performance equivalent to random guessing. 
\textcolor{blue}{
% The same threshold value $\rho$, once determined through this procedure, was applied uniformly across all evaluated models to ensure consistency in the comparison.
To ensure fair and consistent comparison, the same threshold value $\rho$ was uniformly applied across all evaluated models.
}

\subsection{Magnitude evaluation}\label{sec:Magnitude}
To answer RQ$_2$ and evaluate the overlap between the predicted and actual links in our dataset, we adopt three set-based metrics that provide complementary insights into prediction quality. These metrics help us quantify how many relevant links were correctly identified by our model and how well the predicted set matches the actual set.
\textcolor{blue}{In particular, we hypothesize that the best-performing models will reproduce a substantial portion of the ground truth CVE links, although the overlap will remain incomplete due to the inherent limitations of the repositories.}
To test this, we employ three metrics: Jaccard Similarity Index, Mapping Accuracy, and Detection Accuracy.
The Jaccard Similarity evaluates the similarity of two sets: the detected set ($\mathcal{L}_{a}$) and the actual sets  ($\mathcal{M}_{a}$).
It measures the ratio of the intersection of these sets to their union,

\begin{equation}
\mathrm{Jaccard \, Similarity }= \frac{{|\mathcal{L}_{a} \cap \mathcal{M}_{a}|}}{|\mathcal{L}_{a} \cup \mathcal{M}_{a}|}
\end{equation}
\par\noindent
Mapping Accuracy measures the portion of predicted CVE IDs in the set of linked CVE IDs and is the ratio of the intersection of the predicted set ($\mathcal{L}_{a}$) and the actual set ($\mathcal{M}_{a}$) to the size of the actual set ($\mathcal{M}_{a}$).
\begin{equation}
\mathrm{Mapping \, Accuracy} = \frac{{|\mathcal{L}_{a} \cap \mathcal{M}_{a}|}}{{|\mathcal{M}_{a}|}}
\end{equation}
% \enlargethispage{\baselineskip}
Detection Accuracy measures the portion of linked CVE IDs in the set of predicted CVE IDs and is the ratio of the intersection of the predicted set ($\mathcal{L}_{a}$) and the actual set ($\mathcal{M}_{a}$) to the size of the predicted set ($\mathcal{L}_{a}$).
\begin{equation}
\mathrm{Detection \, Accuracy} = \frac{{|\mathcal{L}_{a} \cap \mathcal{M}_{a}|}}{{|\mathcal{L}_{a}|}}
\end{equation}
These metrics provide a comprehensive understanding of the model's ability to accurately detect and map links between the predicted and actual links in the ground truth.

\subsection{Manual validation}\label{sec:Validation}
To answer RQ$_3$, we select the best pair(s) (attack type, model) resulting from the comparison in Section~\ref{sec:Comparison}. We select the same balanced sample we use for the sensitivity analysis in Section~\ref{sec:SensitivityAnalysis} 
\textcolor{blue}{In line with our hypothesis that the model can uncover undocumented but semantically valid attacks to vulnerability links, we analyze the descriptions of the CVEs predicted by the model for these attacks.}
In particular, we focus on all CVE IDs that were predicted but not explicitly linked to the attacks (False Positives). This set includes both the wrongly predicted CVE IDs (i.e., the ``truly" false positives) and the CVE IDs that should have been explicitly linked on the MITRE pages but are instead missing (the imperfect oracle built through the MITRE pages, Section~\ref{sec:AttacksAnnotation}).  
The first and last authors performed a manual iterative validation on the missing CVE IDs.
Missing CVE IDs were considered ``to be linked" when both authors agreed that there was a clear semantic match between the Technique and the CVE descriptions after two iterations.

\section{Results and Discussion}
\label{sec:result}
In this section, we present the results of our experiments by the outcomes for each of the RQs, as follows.

\subsection{\rqone}

To answer RQ$_1$, we evaluate the performance of the 14  sentence transformer models in Table~\ref{tab:ATT2V} over different types of attacks as described in Section~\ref{sec:PerformanceEvaluation}. Specifically, we compute their performance on the description  of (1) Tactic, (2) Technique, (3) Procedure, and (4)  Pattern.  Table~\ref{tab:ATT2V} presents the comparative analysis across all models and types of attacks.
In terms of models, MMPNet achieves the highest F$_1$-score for both Technique (89.0) and Attack Pattern (72.4).
T5 also performs well, achieving the best F$_1$-score in Tactic (88.0) and consistently good results across other attack types with 100\% recall scores, though at the expense of lower precision in some cases (e.g., Procedure: Precision= 48.6).
For what concerns the attack types,
Techniques are the most informative attack type for most models, with many achieving F$_1$-scores above 60\%. MMPNet is the best model for this type.
Procedure is the less informative attack type for the sentence transformers with MSMBERT the best performing model.
\begin{table*}[htbp!]
    \centering
    \fontsize{8pt}{13pt}\selectfont
    \caption{Performance metrics for all types of attacks and models.}
    \begin{tabular}{l@{\hskip 1pt}|c@{\hskip 2pt}c@{\hskip 2pt}c|c@{\hskip 2pt}c@{\hskip 2pt}c|c@{\hskip 2pt}c@{\hskip 2pt}c|c@{\hskip 2pt}c@{\hskip 2pt}c}
    \hline
         &  \multicolumn{3}{c|}{\textbf{Tactic}} & \multicolumn{3}{c|}{\textbf{Technique}} & \multicolumn{3}{c|}{\textbf{Procedure}} & \multicolumn{3}{c}{\textbf{ Pattern}} \\
         \hline 
    \textbf{Acronym} & \textbf{Precision} & \textbf{Recall} & \textbf{F$_1$-Score} & \textbf{Precision} & \textbf{Recall} & \textbf{F$_1$-Score} & \textbf{Precision} & \textbf{Recall} & \textbf{F$_1$-Score} & \textbf{Precision} & \textbf{Recall} & \textbf{F$_1$-Score} \\
    \hline
PAlbert	&	50	&	66.7	&	57.1	&	96.8	&	22.6	&	36.6	&	100	&	6.9	&	12.8	&	47.8	&	50	&	 48.9 \\	
PTinyBERT	&	100	&	18.2	&	30.8	&	100	&	22.6	&	36.8	&	92.1	&	19.1	&	31.6	&	53.3	&	46.5	&	 49.7 \\	  
MDBERT	&	75	&	27.3	&	40	&	100	&	15	&	26.1	&	70.8	&	15.2	&	24.1	&	37.5	&	3.5	&	 6.4 \\	
MSMBERT	&	50	&	100	&	66.7	&	66.5	&	100	&	79.9	&	50	&	100	&	66.7	&	48.6	&	100	&	 65.4 \\	
DRoBERTa	&	66.7	&	18.2	&	28.6	&	79.7	&	41.4	&	54.5	&	77.8	&	8	&	14.4	&	52.1	&	43	&	 47.1 \\	
Roberta	&	75	&	27.3	&	40	&	84.8	&	66.9	&	74.8	&	83.3	&	17	&	28.3	&	52.3	&	53.5	&	 52.9 \\	  \hline
MiniLM6	&	100	&	18.2	&	30.8	&	86.7	&	29.3	&	43.8	&	100	&	1.1	&	2.2	&	48.6	&	20.9	&	 29.3 \\	
MiniLM12	&	100	&	33.3	&	50	&	94.5	&	51.9	&	66.9	&	100	&	3.4	&	6.5	&	52.8	&	32.6	&	 40.3 \\	
MMiniLM6	&	100	&	9.1	&	16.7	&	95.2	&	15	&	26	&	4.4	&	60	&	3.5	&	50	&	10.5	&	 17.3 \\	
PMiniLM6	&	40	&	66.7	&	50	&	93.3	&	31.6	&	47.2	&	86.9	&	2.8	&	5.4	&	52	&	75.6	&	 61.6 \\	
PMiniLM12	&	40	&	66.7	&	50	&	93.7	&	44.4	&	60.2	&	76.9	&	22.7	&	35	&	57.2	&	87.3	&	 69.1 \\	  \hline
MPNet	&	100	&	9.1	&	16.7	&	77.1	&	83.5	&	80.1	&	97.5	&	44.3	&	60.9	&	52.9	&	53.5	&	 53.2 \\	
MMPNet	&	100	&	33.3	&	50	&	84	&	94.7	&	 \textbf{89.0} 	&	89.7	&	29.5	&	44.4	&	71.6	&	73.3	&	 \textbf{72.4}  \\	  \hline
XXLT5	&	78.6	&	100	&	 \textbf{88.0} 	&	66.5	&	100	&	79.9	&	48.6	&	100	&	65.4	&	50.1	&	100	&	 66.7 \\			    \hline
    \end{tabular}
    \label{tab:ATT2V}
\end{table*}
% In contrast, \textit{Procedure} descriptions, which typically refer to fine grained steps with very specific software versions, tools, or configurations}, resulted in the weakest performance across all models. Most models failed to generalize from this type of input, with F$_1$-scores often falling below 50. An exception was the \texttt{msmarco-bert-base-dot-v5} model, which performed comparatively better on this category, achieving a precision of 50.0, a recall of 100, and an F$_1$-score of 66.7. Nonetheless, the high rate of false negatives across other models underscores the challenge of using procedures for broad vulnerability prediction.}
% Furthermore, the \texttt{multi-qa-distilbert-cos-v1}, \texttt{multi-qa-MiniLM-L6-cos-v1}, \texttt{all-MiniLM-L6-v2}, and \texttt{paraphrase-TinyBERT-L6-v2} models performed relatively poorly in comparison for all types of attack descriptions. Specifically, the models failed to achieve the performance metrics for Technique detection with an F$_1$-score of 26.1, 26.0, 43.8, 36.8, respectively. 

\begin{figure*}[htb!]
    \centering
    \includegraphics[width=0.72\textwidth]{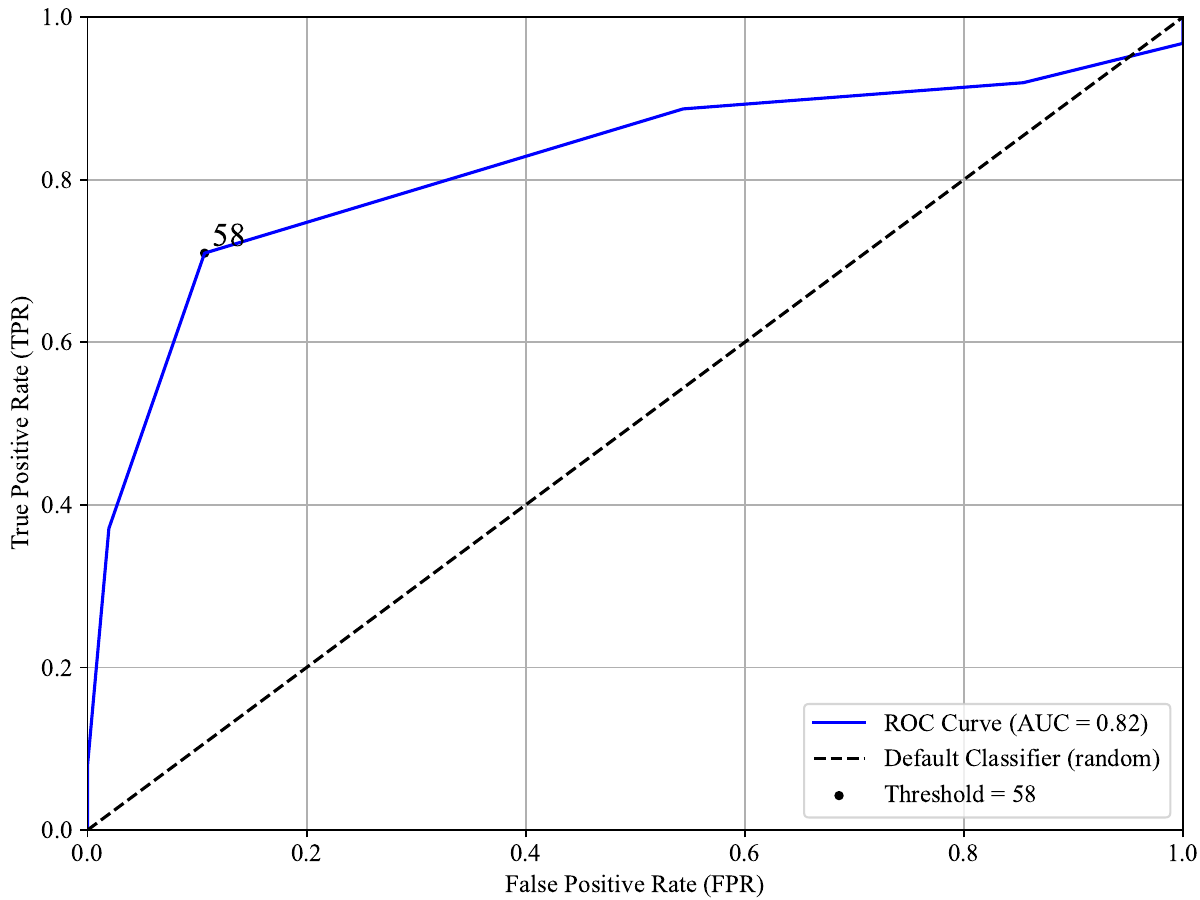}
    \caption{ROC of varying threshold  $\rho$ for pair (MMPNet, Technique). The value nearest to the ideal ratios (0,1) corresponds to $\rho=58$\%.}
    \label{fig:AUCsinsitivity}
\end{figure*}

\textit{Sensitivity analysis.}
As the results in Table~\ref{tab:ATT2V} are computed using a heuristic threshold of $\rho = 60\%$, we further perform a sensitivity analysis (Section~\ref{sec:SensitivityAnalysis}) on the threshold using the best pair (Technique, MMPNet). We perform the analysis on a balanced dataset of Techniques. The dataset comprises 50 positive and 50 negative instances randomly selected from the original dataset. 
Figure~\ref{fig:AUCsinsitivity} shows the ROC curve and the optimal value of the threshold $\rho$. The value of AUC = 0.82 indicates a good overall discriminative ability. 
\textcolor{blue}{In this paper, we used the same similarity threshold $\rho$ value for all models during evaluation.
The selection of the threshold embodies the trade-off between precision and recall: higher recall reduces the likelihood of missing true vulnerabilities, while higher precision decreases false alarms and analyst workload. Although the ROC analysis $\rho = 58$\% confirmed the consistency of our initial choice, we acknowledge that employing a single global threshold across all models and attack types constitutes a limitation of this study in Section~\ref{ch:threatstovalidity}. In practice, individual models or attack types may benefit from tailored thresholds that better capture their specific operating characteristics and optimize the precision–recall balance.}

% \begin{table}[ht]
% \centering
% \caption{Optimal Thresholds for Sentence Transformer Models}
% \begin{tabular}{lc}
% \hline
% \textbf{Model Name} & \textbf{Threshold} \\
% \hline
% multi-qa-mpnet-base-dot-v1 & 58 \\
% paraphrase-multilingual-MiniLM-L12-v2 & 59 \\
% multi-qa-MiniLM-L6-cos-v1 & 57 \\
% multi-qa-distilbert-cos-v1 & 55 \\
% all-MiniLM-L12-v2 & 55 \\
% all-distilroberta-v1 & 57 \\
% all-MiniLM-L6-v2 & 55 \\
% all-mpnet-base-v2 & 56 \\
% paraphrase-MiniLM-L6-v2 & 55 \\
% paraphrase-albert-small-v2 & 55 \\
% msmarco-bert-base-dot-v5 & 56 \\
% all-roberta-large-v1 & 57 \\
% gtr-t5-xxl & 68 \\
% paraphrase-TinyBERT-L6-v2 & 55 \\
% \hline
% \end{tabular}
% \label{tab:thresholds_models}
% \end{table}

\textit{Discussion.}
XXLT5 is a large-scale variant of Google T5  designed for dense retrieval tasks, such as semantic search~\cite{NiEtAl2021}. In our study, it appears to be the most appropriate to detect similarity between Tactics - that have a very abstract description - and CVEs. This model embeds texts in the largest vector space and has the highest number of parameters among all 14 models in Table~\ref{tab:pretrainedmodels}. 
 Such characteristics allow the model to capture more complex patterns and semantic similarities, although the model is more computationally expensive than the others. 
MMPNet is a much thinner and faster model based on the MPNet architecture. Nonetheless, it embeds text in a comparatively large vector space and has a rather large size among the models in Table~\ref{tab:pretrainedmodels}.
\textcolor{blue}{The superior performance of MPNet can be explained by its architecture, which leverages permuted language modeling to better capture bidirectional dependencies and contextual semantics compared to other models.}
In our study, this is the best model for both Techniques and Patterns. This is coherent with the fact that both Techniques and Patterns refer to exploitation methods, Table~\ref{tab:AttacsDescriptions}. 
Interestingly, only MMPNet, together with MSMBERT uses the dot product (instead of cosine) in their loss functions. This allows us to account for the length of the embeddings, not only their angle, and thus they can capture the differences in embeddings based on their context semantics. It is also worth noting that the other dot product model, MSMBERT, is the best model for Procedures. 
\textcolor{blue}{Thus, these results confirm our initial hypothesis for this research question, as Technique descriptions indeed emerged as the most informative attack type, and higher-capacity transformer models demonstrated superior performance compared to lighter architectures.}
%For example, the \texttt{multi-qa-mpnet-base-dot-v1} model has more parameters with a size of 420 MB, which allows it to understand semantic meanings better and work more efficiently. On the other hand, smaller models like the \texttt{multi-qa-distilbert-cos-v1}, \texttt{multi-qa-MiniLM-L6-cos-v1, all-MiniLM-L6-v2, paraphrase-TinyB-ERT-L6-v2}, which are 250, 80, 120, and 240 MB in size, have fewer parameters, which could lower performance. Thus, the F$_1$-scores of the \texttt{multi-qa-mpnet-base-dot-v1} and \texttt{multi-qa-MiniLM-L6-cos-v1} are 89.0 and 26.0, respectively.

\begin{tcolorbox}
{\noindent
\textbf{RQ$_1$ Summary:} Overall, sentence transformers are effective in identifying CVEs from attack descriptions, particularly for Techniques, where the MMPNet model achieves an F$_1$-score of 89. Our study further indicates that the best-performing models either leverage a high-dimensional embedding space with a large number of parameters (e.g., XXLT5), or utilize dot product loss functions, enabling more precise capture of the semantic context between attack and CVE descriptions.
}
\end{tcolorbox}

\subsection{\rqtwo}

\begin{figure}[htb!]
\centering
\includegraphics[width=\columnwidth]{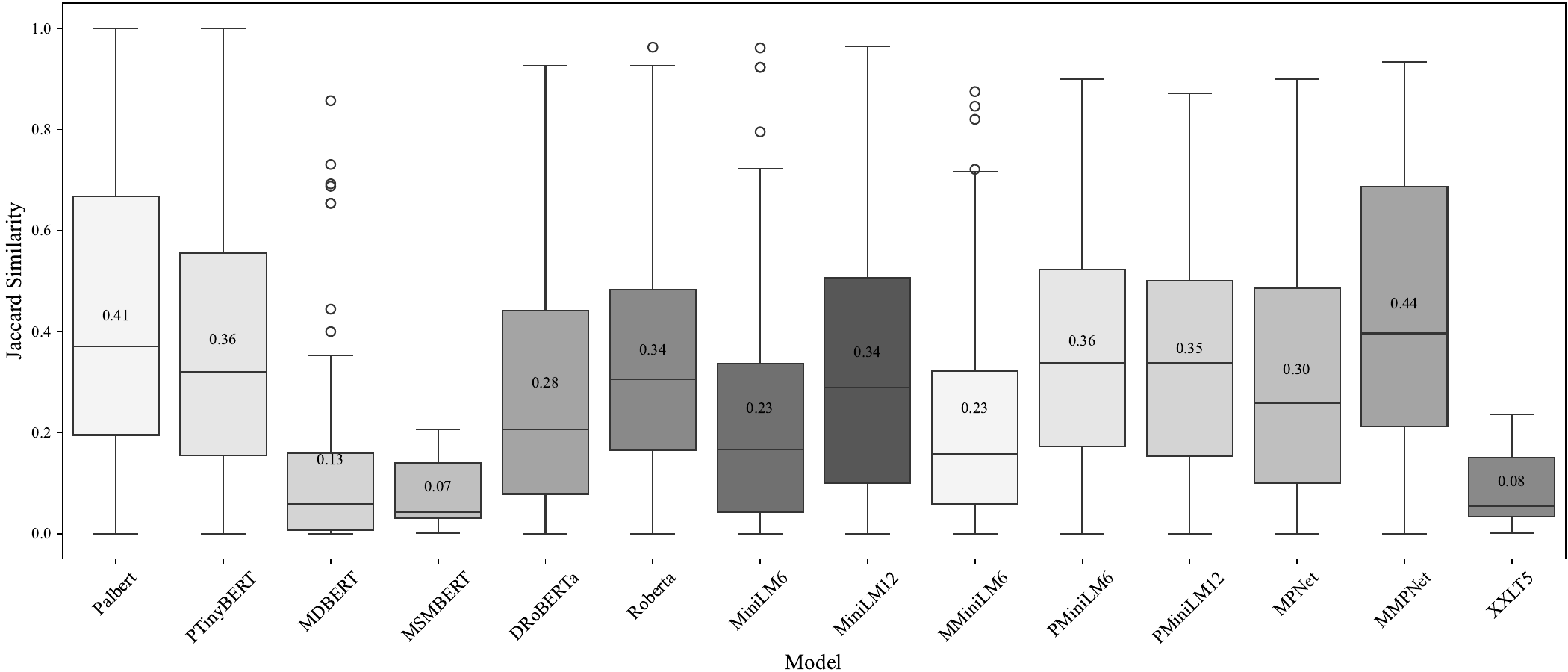}
% \captionsetup{font=small} % Set the font size of the caption to small
\caption{Jaccard similarity across SOTA models.}\label{fig:allmodelsJaccard}
%\vspace{-.3cm}
\end{figure}

\textcolor{blue}{
To answer RQ$_2$, we evaluate, for each Technique ID, how many of the predicted CVE IDs are actually linked. This evaluation began with a comparative analysis of SOTA transformer-based models, using Jaccard Similarity as the primary metric (Figure~\ref{fig:allmodelsJaccard}). The figure indicates that MMPNet achieved the highest Jaccard Similarity (mean = 0.44), demonstrating superior capability in identifying linked CVE IDs from attack descriptions. Based on this observation, MMPNet was selected for further assessment using three complementary metrics, Jaccard Similarity, Mapping Accuracy, and Detection Accuracy, as defined in Section~\ref{sec:Magnitude}.
}
% To answer RQ$_2$, we evaluate, for each Technique ID, how many of the predicted CVE IDs are actually linked, using the MMPNet model. We evaluate the model’s output using three  metrics, Jaccard Similarity, Mapping Accuracy, and Detection Accuracy defined in Section~\ref{sec:Magnitude}. 

\begin{figure}[htb!]
\centering
\includegraphics[width=.65\columnwidth]{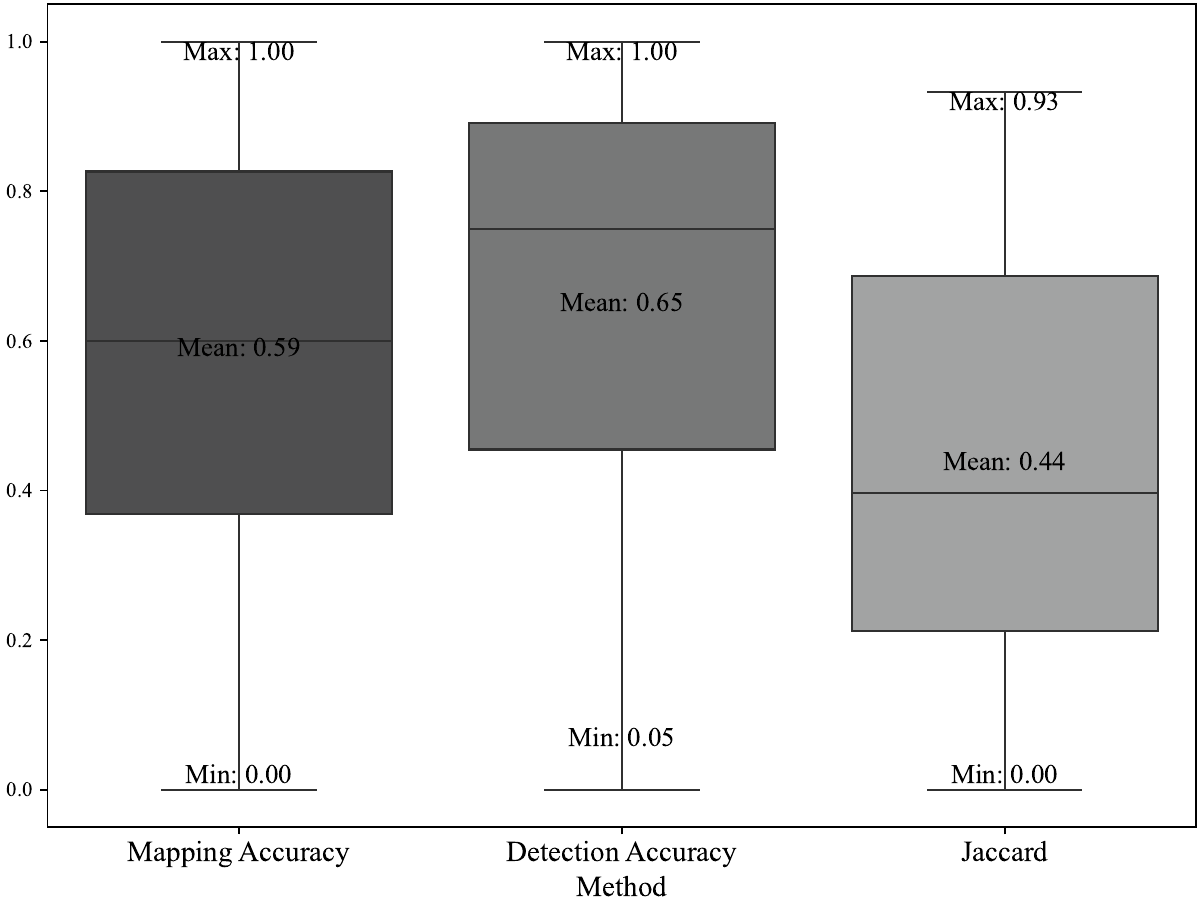}
% \captionsetup{font=small} % Set the font size of the caption to small
\caption{Performance of the MMPNet-based approach across attack techniques description.}\label{fig:jaccardplot}
%\vspace{-.3cm}
\end{figure}
Figure~\ref{fig:jaccardplot} reports the distribution of these measures across all Techniques.
On average, the Mapping and Detection accuracies report more than random guessing (50\%), but also report a large amount of undetected linked CVE IDs (44\% on average) and a good amount of predicted unlinked CVE IDs (39\%).
For more than 50\% of the Techniques the model predicts CVE IDs that are over 70\%  linked (i.e., a median Detection Accuracy above 70\%). Conversely, for more than 50\% of Techniques, over 50\% of their linked CVE IDs are correctly predicted.
 A representative example is  Technique T1539 ``Steal Web Session Cookie" that achieves the highest Jaccard Similarity score of 82\% obtained as the ratio between the 124 CVE IDs in $\mathcal{L}(T1539) \cap \mathcal{M}(T1539)$ and the 151 in $\mathcal{L}(T1539) \cup \mathcal{M}(T1539)$. As $\mathcal{L}$ contains 150 CVE IDs and $\mathcal{M}(T1539)$ contains 125 CVE IDs, the capability of MMPNet is very high (82\% and 99\%). The model finds all but one existing linked CVE IDs and a good number of additional ones. 
The additional predicted CVE IDs may be false positives as well as potential missing CVE IDs, i.e. links that may not yet be documented.
% On the other hand,  a lower Jaccard Similarity value suggests limited agreement between the elements of the real and detected sets.  Specifically,  T1003 Technique achieved a Jaccard Similarity of only 5\%.}
%  Upon manual evaluation of this attack by examining which CVE are related to the attack Technique text, our approach discovered the size of   $\mathcal{L}_{T1003}$ is 37, and we found 22 of these links do not exist in the MITRE repositories.
% Thus, it is important to note that many detected CVE links may indeed be linked to the attack Technique. However, their link is missing from the MITRE repositories.
The next research question aims to address the issue of the imperfect oracle and the potential missing links. 
\textcolor{blue}{Thus, these findings confirm our hypothesis for this research question, since the best-performing models reproduced a significant portion of existing CVE links while leaving space for additional, potentially missing ones.}

\begin{tcolorbox}
{\noindent
\textbf{RQ$_2$ Summary.} The overlapping between predicted and linked CVE IDs is satisfactory, reaching a Mapping Accuracy of 56\%, Detection Accuracy of 61\%, and Jaccard Similarity of 40\%. Inspecting the metrics for representative cases we noticed that some predicted CVEs may be missing in the MITRE pages and can eventually be added.
% , indicating many predicted CVEs links may be linked to Attack Patterns but are not present in the MITRE repositories
}
\end{tcolorbox}

\subsection{\rqthree}
% The results were discussed at the end of each run, and an agreement was reached. 

\begin{table*}[htb]
\centering
\footnotesize
% \scriptsize % Optionally reduce font size further
\caption{Summary of our approach validation process for new candidates' links.}
\label{tab:validationResult}
% Adjusting column spacing and row height
\setlength{\tabcolsep}{3pt} % Adjust as needed
\begin{tabular}{@{}lcc|cc@{}} % Removing padding around columns
\hline
&\textbf{Retrieved}&&\textbf{Validated}&\\
\hline
&\textbf{Total Predicted}&\textbf{False positives}&\textbf{False recommended}& \textbf{Recommended to be linked} \\
\hline
% \textbf{Total CVE Links} &  6453 &605&467& 138\\
\textbf{Number of linked CVE reports} &  2230 &434& 159& 275
 (12.3\%)\\
\textbf{Number of techniques} & 201 & 74 & 54 & 60 (29.8\%)\\
 \hline
\end{tabular}
\end{table*}
\textcolor{blue}{
Table~\ref{tab:validationResult} presents the CVEs predicted by our model for a selected subset of 100 technique descriptions.} We summarize the unlinked Techniques predicted with the model MMPNet. For 201 Techniques we found 434 predicted and unlinked CVE IDs. Among the latter, after the manual inspection (Section~\ref{sec:Validation}),  159 were confirmed as falsely predicted and the remaining 275 predicted by 60 Techniques as to be linked~\footnote{It is worth noticing here that one CVE report can be linked to more than one Technique. Thus, the Technique numbers in the last two columns do not sum up to the value of the second column}.  The full list of the 275 manually validated links is available as supplementary material at GitHub~\cite{VULDAT}.
As an illustrative example, we discuss the recent story of Technique T1039 ``Data from Network Shared Drive" and the  CVEs our approach predicted, namely CVE-2005-1205, CVE-2009-3107, and CVE-2020-3452. All these vulnerabilities involve unauthorized access or data exposure through shared network resources, which aligns well with the description of T1039. Although these links are now included in the ATT\&CK repositories, they were missing when we annotated the attack datasets, as in Section~\ref{sec:AttacksAnnotation}. In other words, we were able to uncover missing CVE IDs that were independently added to the MITRE repository later on. This further reinforces the validity of our approach. 
We have already established contact with members of the MITRE board and are in the process of preparing a formal submission of our manually validated links to support the enrichment of MITRE repositories.
\textcolor{blue}{These results show that our approach successfully identified 275 previously undocumented attacks to vulnerability links across 60 Techniques, many of which were later verified or added to the MITRE repositories. This confirms our hypothesis for RQ$_3$, as it demonstrates that the current repositories are incomplete and can indeed be enriched through automated linking.}
\begin{tcolorbox}
{\noindent
\textbf{RQ$_3$ Summary}: Our approach is able to recommend 275 missing links across 60 Techniques, highlighting its potential to enhance the existing information in the MITRE repositories and increase coverage by over 12\%.
% By recommending 275 missing CVE links across 60 attack techniques, our approach demonstrates its potential to enhance existing repositories and expand coverage by more than 13\%.
% , corresponding to about 13\% of the predicted links and 30\% of all techniques.

}
\end{tcolorbox}

\subsection{Implications for practice and research}
\label{Sec:implication}
% \textcolor{blue}{
% This study demonstrates the utility of transformer-based language models for attack-to-CVE linking. We evaluate 14 SOTA sentence transformers and compare four attack description types: Tactic, Technique, Procedure, and Attack Pattern, within a unified pipeline. We further report 275 undocumented Technique–CVE links and introduce a manual validation stage for candidates absent from MITRE to reduce false positives.}

\textcolor{blue}{This study evaluates 14 SOTA transformers across four attack description types (i.e., \textit{Tactic}, \textit{Technique}, \textit{Procedure}, and \textit{Attack Pattern}). We show that Technique descriptions yield the highest predictive performance (MMPNet: F$_1{=}89$), and we report \textit{275} previously undocumented Technique–CVE links. Taken together, these results provide empirical evidence that \textit{manual maintenance and independently evolving repositories are insufficient to achieve timely and comprehensive cross-referencing at this scale}, thereby motivating automation.
By coupling comparative model evidence with a curated set of 275 new links, this work provides both a practical pathway for decision-makers to accelerate mitigation and a research scaffold for advancing scalable, transparent, and operationally relevant attack–vulnerability linking. The following details the main implications for both practice and research fields.} 
\textcolor{blue}{
\paragraph{Implications for practice}
The evidence provides actionable guidance for security operations and vulnerability management:
(i) prioritize Technique descriptions when predicting candidate CVEs, since they carry the strongest signal for linking; 
(ii) operationalize high-performing models (e.g., MMPNet) to pre-populate candidate CVE sets for each observed technique, thereby reducing analyst time on initial triage; 
and (iii) integrate the \textit{275 validated new links} into vulnerability management systems and threat intelligence platforms to enrich lookups and shorten patching lead times; Collectively, these practices support earlier mitigation, more consistent triage, and stronger defensive posture.
}
% \textcolor{blue}{
% For researchers, the findings from this study and the proposed multi type attack–to–vulnerability linking framework open several promising avenues for further investigation and advancement in the field of automated vulnerability
% detection. One potential avenue is to extend the evaluation of the 14 sentence transformer models to additional transformer architectures or domain-specific language models, applying the approach to diverse cyber threat datasets and multilingual attack reports. This could involve testing the approach's adaptability across broader contexts and refining its performance in varied operational environments. Another valuable direction is advancing the approach toward real-time applications that process attack news from cybersecurity magazines
% would enable the immediate identification and recommendation of relevant CVEs across multiple languages and data
% sources, improving global threat awareness and response speed. Incorporating continuous data streams from multiple
% intelligence feeds, retraining models to adapt to emerging vulnerabilities, and developing interpretability and visualization tools for model outputs are also important areas for future research. Collectively, these directions provide a foundation for scalable, transparent, and operationally relevant vulnerability–attack linking systems that can both advance academic understanding and enhance real-world cybersecurity practices.}

\textcolor{blue}{
\paragraph{Implications for research}
The findings from this study and the comparative evaluation across 14 transformers and four attack information types establish a replicable baseline for automated attack–to–vulnerability linking. Building on this baseline, promising directions include:
(i) model evolution, assessing newer architectures and domain-specific or multilingual models for robustness across heterogeneous cyber threat datasets. This could involve testing the approach’s adaptability across broader contexts and refining its performance in varied operational environments; 
(ii) data breadth, extending beyond ATT\&CK–CVE to additional feeds and assessing generalization under distribution shift; 
(iii) timeliness, moving toward near real-time ingestion of attack reports (news, advisories) to recommend relevant CVEs as events unfold, improving global threat awareness and response speed; and 
(iv) transparency, incorporating continuous data streams from multiple intelligence feeds, retraining models to adapt to emerging vulnerabilities, and developing interpretability and visualization tools for model outputs are also important areas for future research. Collectively, these directions provide a foundation for scalable, transparent, and operationally relevant vulnerability–attack linking systems that can both advance academic understanding and enhance real-world cybersecurity practices.}
\section{Threats to validity}
\label{ch:threatstovalidity}
In this section, we discuss the potential threats to the validity of our study in terms of construct, internal, and external validity.
\par\noindent
\textbf{Construct validity.}
Construct validity refers to how much the results at the operational level support the claims at the conceptual level~\cite{sjoberg2022construct}. 
In our case, two main construct threats arise: the choice of cosine similarity threshold ($\rho$) and the input size for transformer models. 
The similarity threshold $\rho$ is not absolute. We empirically determined the value that balances false positives and false negatives, but different values can have an impact on the results, knowing that a larger $\rho$ results in a more conservative approach in determining true values, but leading to a lower recall. In contrast, a smaller $\rho$ may yield a more optimistic outcome when determining true values, increasing recall, but at the expense of precision. \textcolor{blue}{To validate our choice, we employed ROC analysis, which confirmed the robustness of the selected threshold. Nevertheless, using a single global threshold across all models and attack types may not be optimal, as individual models could benefit from tailored thresholds that better capture their operating characteristics.}
Additionally, transformer models are constrained to processing a fixed number of input tokens,
the first 384 words, which could potentially lead to context truncation and bias toward initial content. Nevertheless, this limitation does not affect our study, as the attack and vulnerability descriptions used are all within the 384-token limit, not impacting the connection to our conceptual representation of semantic links.

% Figure.~\ref{fig:techwordcount} illustrates the word count of the technique descriptions analyzed in our study.
% \begin{figure*}[htb!]
% \centerline{\includegraphics[width=\textwidth]{Fig/TechWordCount.pdf}}
% \caption{Number of words of techniques description.}
% \label{fig:techwordcount}
% \end{figure*}

\par\noindent
\textbf{Internal validity.}
Internal validity refers to the extent to which treatment and outcome variables are related to the causal relation between the treatment and outcome constructs~\cite{sjoberg2022construct,Shadish2002expdesign}. 
In this work, the attack and vulnerability texts may contain technical terms or acronyms (e.g., software versions or specific library names), which may introduce noise in the semantic representation and influence model performance.
% that may hamper the learning and affect the study's internal validity. 
To mitigate this, we applied consistent pre-processing steps to clean and normalize the input texts.
% We have implemented a careful pre-processing to control and, eventually, remove such tokens.
Another potential threat arises from our use of MITRE repositories to construct the ground truth, which may evolve over time. To reduce this impact, we used the most recent version of each repository available at the time of the study and documented our data snapshot.

\par\noindent
\textbf{External validity.}
% When discussing the generalization of results across different datasets, external validity refers to the ability to transfer the results gained from a study on one dataset to a different dataset. 
% Our analysis faces two external threats in this work: the connectivity between attacks and vulnerabilities and the potential for the ATT\&CK framework we utilized not to be consistently updated.
% The links between attacks and vulnerabilities we retrieved from the MITRE repositories served as the study's ground truth, but their number might be limited. 
% Out of the 201 techniques listed in the ATT\&CK framework and the 295,604 CVE reports in the CVE repository, our ground truth is built only on links between  100 Subtechniques and 610 CVE reports. 
% Nonetheless, this information shows that it suffices for the exploratory study we have performed.  
% The attack techniques and CVE descriptions may change over time. Thus, the dataset we rely on may change as well. Furthermore, the ATT\&CK framework used in our study may not always be up-to-date or comprehensive with information about attacks and vulnerabilities.  In the future, we will explore other sources (e.g., the exploit-DB repository~\cite{exploitdb}).
If there is a causal relationship between the construct and the effect, external validity tells us that the results can be generalized outside the scope of the study~\cite{wohlin2012experimentation}. In our case, external validity concerns mainly the generalizability of our findings to other contexts or datasets. This study faces two external threats: First, the ground truth was constructed using a subset of 100 ATT\&CK subtechniques linked to 610 CVE reports. While this sample is suitable for the purposes of our exploratory analysis, it does not reflect the entire scope of the MITRE repositories, which include over 201 techniques and more than 295,000 CVEs. 
Second, both the ATT\&CK and CVE repositories are continuously updated, meaning that the structure, content, and relationships within the data may evolve over time. As a result, models trained on a static snapshot may not maintain the same level of performance on future data. To enhance the generalizability of our approach, future work will explore integrating additional sources such as ExploitDB~\cite{exploitdb} and will assess the approach on other publicly available threat intelligence datasets.

\section{Related work}
\label{ch:relatedwork}
In this section, we review the related work on vulnerability attack models. 
Existing studies are grouped into two categories: vulnerability-to-attack mapping and attack-to-vulnerability mapping.
\par\noindent
 \textbf{Vulnerability-to-Attack Mapping:} The majority of research studies have focused on linking vulnerabilities to attacks and not the other way around. The multi-head deep embedding model~\cite{kuppa2021linking} learns links between CVE reports and ATT\&CK techniques. Using regular expressions in the attack reports. Then, it compares the ATT\&CK Technique vectors with the text description found in the CVE metadata using the cosine distance. This study only maps 17 ATT\&CK  techniques.
Researchers have used the BERT pre-trained transformer model~\cite{devlin2018bert} to extract information from the vulnerability database to improve the textual descriptions of newly discovered CVE reports~\cite{sun2021generating}.
A multi-label classification approach ~\cite{lakhdhar2021machine} introduced for automatically mapping CVE reports to ATT\&CK tactics in the MITRE repositories evaluated several machine learning algorithms to find the best approach.
Another model mapping between vulnerabilities and attacks is Cve2att\&ck ~\cite{grigorescu2022cve2att}. The model annotates a dataset of CVE reports with ATT\&CK  tactics, using BERT-based language models and classic models for machine learning.  The study is limited to a set of 31 ATT\&CK techniques and a training set of 1813 CVE reports. 
The CVE Transformer (CVET) ~\cite{ampel2021linking} is a model that combines a self-awareness distillation design utilized for fine-tuning the pre-trained language model RoBERTa ~\cite{liu2019roberta}. The main objective is linking a CVE to one of 10 ATT\&CK tactics.
\begin{table*}[htbp!]
    \centering
    \scriptsize
    \caption{Summary of Related Work in Vulnerability-Attack Mapping}
    \label{tab:related_work}
    \begin{tabular}{l@{\hskip 4pt}p{0.18\textwidth}@{\hskip 4pt}p{0.3\textwidth}@{\hskip 4pt}p{0.25\textwidth}}
    \hline
        \textbf{Study} & \textbf{Approach}& \textbf{Key Contribution} & \textbf{Limitations} \\
    \hline
    \textbf{Kuppa, A. et al.~\cite{hemberg2022sourcing}}&Multi-head Deep: Cosine similarity on embeddings.& Links CVEs to 17 ATT\&CK techniques using ATT\&CK and CVE datasets. & Limited scope, regex dependency. \\
    \textbf{Grigorescu, Octavian, et al.~\cite{grigorescu2022cve2att}} & Cve2att\&ck: BERT-based classification. &Annotates CVEs with tactics using 1,813 CVEs and 31 ATT\&CK techniques. & Small training set, narrow Technique coverage. \\
    \textbf{Ampel, Benjamin, et al.~\cite{ampel2021linking}} & CVET: RoBERTa fine-tuning. &Self-awareness distillation for Tactic mapping using 10 ATT\&CK tactics. & Only 10 tactics, no technique-level granularity. \\
    \textbf{Kanakogi, Kenta, et al..~\cite{kanakogi2021tracing}}&CAPEC-CVE: TF-IDF, RoBERTa, Doc2Vec. &Top-$N$ CAPEC matches for CVEs using CAPEC and CVE datasets. & Unidirectional (CVE$\to$CAPEC). \\
    \textbf{Othman et al.~\cite{othman2024comparison}} & CAPEC-CVE & Empirical comparison models (TF-IDF, LSI, BERT, MiniLM, RoBERTa)  and provides a mapping dataset linking 133 CAPEC attack patterns to 685 CVEs via CWEs.&  TF-IDF and LSI fail to capture contextual or semantic relationships in attack descriptions. \\
    \textbf{Hemberg, Erik, et al.~\cite{hemberg2021linking}} &  BRON: graph aggregation. &Bidirectional relational path tracing using CWE, CVE, CAPEC, and ATT\&CK datasets. & Fails on recent CVEs, no NLP integration. \\
   
\textbf{Othman et al.~\cite{othman2024cybersecurity}} & Technique-CVE. & Empirical comparison transformers models (BERT, MPNeT, MiniLM)  and provides a mapping dataset linking 100 attack techniques to 610 CVEs via CWEs and CAPEC.& Limited in its ability to compare all types of attack information (Tactic, Technique, Procedure, and Attack Pattern).\\
   \textcolor{blue}{
\textbf{Othman et al. (This study)}} & \textcolor{blue}{Att\&ck2Cve: Transformer-based approach for linking attacks to CVEs across four attack information types (Tactic, Technique, Procedure, Attack Pattern) using cosine similarity on embeddings.}
&
\textcolor{blue}{
Evaluation of 14 SOTA sentence transformers, accompanied by a comprehensive analysis that identifies the most effective attack information type (Technique) for vulnerability prediction; the MMPNet model achieves an F$_1$-score of 89\% and enables the identification of 275 validated missing CVE links, thereby enriching the MITRE repositories.}
&
\textcolor{blue}{
Although the limitations of related work have been addressed, we acknowledge that further improvements remain possible, as discussed in Section 5.}
 \\

    \hline
    \end{tabular}
\end{table*}
\par\noindent
\textbf{Attack-to-vulnerability mapping:} The majority of research studies have focused on utilizing attack patterns in CAPEC rather than leveraging TTP information from the MITRE repositories.
Based on the similarity between the CAPEC document and the CVE description, NLP-based techniques are also utilized to establish the link between CAPEC documents and  CVE reports~\cite{kanakogi2021tracing, hemberg2022sourcing} with RoBERTa outperforming the general BERT model. Other NLP techniques are used to build a direct correlation between CVE and CAPEC reports, such as TF-IDF ~\cite{ramos2003using} and Doc2Vec ~\cite{lau2016empirical}. The optimal link is provided by the TF-IDF when using CAPEC information and CVE descriptions, also providing the \textit{top-n} CAPEC documents that match the CVE description~\cite{othman2024comparison, kanakogi2022comparative}. 
To further improve attack-to-vulnerability mapping, we developed VULDAT~\cite{othman2024cybersecurity}, an automated tool leveraging the MPNet sentence transformer to link attack technique descriptions to CVE vulnerabilities. Unlike previous approaches that focused solely on CAPEC patterns, VULDAT exclusively utilizes attack Technique descriptions to establish links with vulnerabilities. 
% In this study, we expand on this work by evaluating 14 SOTA sentence transformers across multiple attack description types to determine the most effective type for automated vulnerability identification.
BRON ~\cite{hemberg2021linking} is a bi-directional aggregated data graph that supports relational path tracing between CWE, CVE, CAPEC, and ATT\&CK tactics and techniques. Through investigations of the resulting graph representation and data mining of the relational connections between all these cybersecurity knowledge sources, BRON builds a graph framework that promises to integrate all scattered data. However,  the model fails because it cannot link the most recent CVE reports to ATT\&CK Enterprise Matrix techniques.

\textcolor{blue}{
A comparison of representative prior works is summarized in Table~\ref{tab:related_work}. Previous studies have primarily focused on linking CVEs to a single type of attack information, such as Tactics or Techniques~\cite{hemberg2022sourcing,ampel2021linking,grigorescu2022cve2att}. For example, CVET~\cite{ampel2021linking} achieved an F$_1$-score of 76.2\% when mapping CVEs to ATT\&CK Tactics, while Cve2att\&ck~\cite{grigorescu2022cve2att} reported an F$_1$-score of 47.8\% for CVE-to-Technique mappings. Similarly, Kanakogi et al.~\cite{kanakogi2021tracing} achieved an F$_1$-score of 44.5\% when linking CVEs to CAPEC attack patterns. Our earlier work, VULDAT~\cite{othman2024cybersecurity}, attained an F$_1$-score of 85\% for linking Technique descriptions to CVEs.
In this paper, we conduct a comprehensive evaluation of 14 SOTA sentence transformer models. Additionally, we assess the performance of these models across four distinct types of attack descriptions: Tactic, Technique, Procedure, and Attack Pattern, providing a systematic comparison of model effectiveness. 
Our evaluation reports that Technique descriptions are the most informative for vulnerability prediction and that MMPNet achieves an F1-score of 89\%. Furthermore, our method in this article also incorporates a manual validation stage, leading to the identification of 275 undocumented Technique-CVE links, thereby enriching the MITRE repositories and enhancing their utility for cyber threat intelligence.
}

% A comparison of representative prior works is summarized in Table~\ref{tab:related_work}. Previous studies have primarily focused on linking CVEs to a limited attack information, such as tactics or techniques~\cite{hemberg2022sourcing,ampel2021linking,grigorescu2022cve2att}. For instance, CVET~\cite{ampel2021linking} achieved an F$_1$-score of 76.2\% when mapping CVEs to ATT\&CK tactics, while Cve2att\&ck~\cite{grigorescu2022cve2att} reported a much lower F$_1$-score of 47.8\% for CVE-to-Technique mappings. Similarly, Kanakogi et al.~\cite{kanakogi2021tracing} reported a 44.5\% F$_1$-score when linking CVEs to CAPEC attack patterns. In contrast, our earlier work, VULDAT~\cite{othman2024cybersecurity} achieved a substantially higher F$_1$-score of 85\% in linking Technique descriptions to CVEs.In this paper, we extend beyond prior efforts in several key aspects. First, we conduct a comprehensive analysis of 14 SOTA sentence transformer models. Additionally, we evaluate the performance of these models across four distinct types of attack descriptions: Tactics, Techniques, Procedures, and Attack Patterns, providing deeper insight into which attack information is most effective for vulnerability prediction. Furthermore, our approach supports analyzing both known and potentially new links between attacks and vulnerabilities, offering a more complete and automated framework for vulnerability detection.

\section{Conclusion and Future work}
\label{ch:conclusion} 
In this study, we introduced a novel approach utilizing 14 sentence transformer models to automatically identify CVE reports from textual descriptions of adversary behaviors, also known as attacks. Our approach relies on the  MITRE family of repositories, which represents a comprehensive knowledge base of adversary tactics and techniques as input information for CVE prediction. We constructed an annotated dataset leveraging the information contained in such repositories. We further conducted an experiment to assess the performance of each model using different types of information about an attack stored in the MITRE repositories. The results indicate that our approach performs the best when using the Technique descriptions as input, achieving an F$_1$-score of 89\%. Our approach can supply a valuable service to the cybersecurity community by providing the first approach to automatically link attacks to vulnerabilities and their mitigation strategies, rather than the other way around. Our manual validation of our approach discovered 275 links between attacks and vulnerabilities not contained in the MITRE repositories. These correspond to 12.3\% of the total links found by our model.

% As such, VULDAT can be incorporated into development pipelines for security checks.  
% Our manual validation of VULDAT output discovered 275 links between attacks and vulnerabilities not contained in the MITRE repositories. These correspond to 12.3\% of the total links found by VULDAT. 

%We enhanced our dataset through a thorough validation procedure, making it more beneficial for security research and analysis by demonstrating that VULDAT can recommend missing links between attack techniques and vulnerabilities with a 1.2\% recommendation rate. This could allow VULDAT to recommend new links not yet connected in repositories.
% We propose potential future directions for related work in this domain.
% Additionally, we will explore how these models can predict CVE issues from publicly available cyberattack news articles, such as those published by online cybersecurity magazines (e.g., SecurityWeek) and vendor threat reports. These sources typically describe real-world incidents and emerging threats in unstructured text, providing valuable context for vulnerability detection.
We plan to evaluate the performance of transformer models on other annotated datasets. Additionally, we will explore how these models can predict CVE issues from publicly available cyberattack news reports, such as those published by online cybersecurity magazines (e.g., SecurityWeek~\cite{SecurityWeek}). These sources typically describe real-world incidents and emerging threats in unstructured text, providing valuable context for vulnerability detection and enabling the early identification of vulnerabilities before they are exploited.
Furthermore, we plan to extend our approach by conducting a study that establishes links between attacks, vulnerabilities, and weaknesses in code. Additionally, we aim to identify and provide techniques that can be used to exploit these weaknesses in code.
Finally, we have already begun engaging with the MITRE board to discuss integrating our newly identified links into their existing listings. 

\section{Acknowledgements}
The authors extend their gratitude to the CSLab at the Free University of Bozen-Bolzano for supporting this work under project number EFRE1039 as part of the EFRE-FESR 2021-2027 program.

 \bibliographystyle{elsarticle-num} 
 \bibliography{References}

\begin{thebibliography}{10}
\expandafter\ifx\csname url\endcsname\relax
  \def\url#1{\texttt{#1}}\fi
\expandafter\ifx\csname urlprefix\endcsname\relax\def\urlprefix{URL }\fi
\expandafter\ifx\csname href\endcsname\relax
  \def\href#1#2{#2} \def\path#1{#1}\fi

\bibitem{admass2024cyber}
W.~S. Admass, Y.~Y. Munaye, A.~A. Diro, Cyber security: State of the art, challenges and future directions, Cyber Security and Applications 2 (2024) 100031.

\bibitem{Cybercrime}
M.~M. Robert~Muggah, Cybercrime to cost the world 10.5 trillion annually by 2025, accessed: January 28, 2024. \url{ https://www.weforum.org/agenda/2023/01/global-rules-crack-down-cybercrime/} (2023).

\bibitem{CResearch}
C.~Point, 38\% increase in 2022 global cyberattacks, \url{https://blog.checkpoint.com/2023/01/05/38-increase-in-2022-global-cyberattacks/} (2024).

\bibitem{rahman2023attackers}
M.~R. Rahman, R.~M. Hezaveh, L.~Williams, What are the attackers doing now? automating cyberthreat intelligence extraction from text on pace with the changing threat landscape: A survey, ACM Computing Surveys 55~(12) (2023) 1--36.

\bibitem{ATTACK}
MITRE, Attack, \url{https://attack.mitre.org/} (2025).

\bibitem{CAPEC}
MITRE, Capec, \url{https://capec.mitre.org/} (2025).

\bibitem{CWE}
MITRE, Cwe dataset, \url{https://cwe.mitre.org/} (2025).

\bibitem{CVEdataset}
MITRE, Cve, \url{https://www.cve.org} (2024).

\bibitem{refat2024comparison}
O.~Refat, R.~Bruno, R.~Barbara, A comparison of vulnerability feature extraction methods from textual attack patterns, in: 2024 50th Euromicro Conference on Software Engineering and Advanced Applications (SEAA), IEEE, 2024.

\bibitem{WhatCVE}
T.~Armerding, Cve definitions, \url{https://www.csoonline.com/article/3204884/what-is-cve-its-definition-and-purpose.html}.

\bibitem{othman2024vulnerability}
R.~T. Othman, Vulnerability detection for software-intensive system, in: Proceedings of the 28th International Conference on Evaluation and Assessment in Software Engineering, 2024, pp. 510--515.

\bibitem{sonmez2021classifying}
F.~{\"O}. S{\"o}nmez, Classifying common vulnerabilities and exposures database using text mining and graph theoretical analysis, Machine Intelligence and Big Data Analytics for Cybersecurity Applications (2021) 313--338.

\bibitem{elder2022really}
S.~Elder, N.~Zahan, R.~Shu, M.~Metro, V.~Kozarev, T.~Menzies, L.~Williams, Do i really need all this work to find vulnerabilities? an empirical case study comparing vulnerability detection techniques on a java application, Empirical Software Engineering 27~(6) (2022) 154.

\bibitem{othman2024cybersecurity}
R.~Othman, B.~Rossi, B.~Russo, Cybersecurity defenses: Exploration of cve types through attack descriptions, in: 2024 50th Euromicro Conference on Software Engineering and Advanced Applications (SEAA), IEEE, 2024, pp. 415--418.

\bibitem{VULDAT}
R.~Othman, Att\&ck2vul - automated vulnerability detection from cyberattack text, accessed: Feb 2, 2025. \url{https://github.com/ref3t/Attack2VUL/tree/main} (2025).

\bibitem{reimers2019sentence}
N.~Reimers, I.~Gurevych, Sentence-bert: Sentence embeddings using siamese bert-networks, in: Proceedings of the 2019 Conference on Empirical Methods in Natural Language Processing and the 9th International Joint Conference on Natural Language Processing (EMNLP-IJCNLP), Association for Computational Linguistics, 2019.

\bibitem{raffel2020exploring}
C.~Raffel, N.~Shazeer, A.~Roberts, K.~Lee, S.~Narang, M.~Matena, Y.~Zhou, W.~Li, P.~J. Liu, Exploring the limits of transfer learning with a unified text-to-text transformer, Journal of machine learning research 21~(140) (2020) 1--67.

\bibitem{VULDATDataSet}
R.~Othman, Vuldat- vulnerability dataset, accessed: Feb 2, 2025. \url{ figshare. Dataset. https://doi.org/10.6084/m9.figshare.25828102.v1} (2025).

\bibitem{dong2023dekedver}
Y.~Dong, Y.~Tang, X.~Cheng, Y.~Yang, Dekedver: A deep learning-based multi-type software vulnerability classification framework using vulnerability description and source code, Information and Software Technology 163 (2023) 107290.

\bibitem{elder2024survey}
S.~Elder, M.~R. Rahman, G.~Fringer, K.~Kapoor, L.~Williams, A survey on software vulnerability exploitability assessment, ACM Computing Surveys 56~(8) (2024) 1--41.

\bibitem{dempsey2017automation}
K.~Dempsey, P.~Eavy, G.~Moore, Automation support for security control assessments, Vol. 1: Overview (2017) 8011--1.

\bibitem{esposito2024validate}
M.~Esposito, D.~Falessi, Validate: A deep dive into vulnerability prediction datasets, Information and Software Technology (2024) 107448.

\bibitem{alevizopoulou2021social}
S.~Alevizopoulou, P.~Koloveas, C.~Tryfonopoulos, P.~Raftopoulou, Social media monitoring for iot cyber-threats, in: 2021 IEEE International Conference on Cyber Security and Resilience (CSR), IEEE, 2021, pp. 436--441.

\bibitem{gasmi2019information}
H.~Gasmi, J.~Laval, A.~Bouras, Information extraction of cybersecurity concepts: An lstm approach, Applied Sciences 9~(19) (2019) 3945.

\bibitem{vuldatPaper}
R.~Othman, B.~Russo, Vuldat: Automated vulnerability detection from cyberattack text, in: C.~Silvano, C.~Pilato, M.~Reichenbach (Eds.), Embedded Computer Systems: Architectures, Modeling, and Simulation, Springer Nature Switzerland, Cham, 2023, pp. 494--501.

\bibitem{theisen2018attack}
C.~Theisen, N.~Munaiah, M.~Al-Zyoud, J.~C. Carver, A.~Meneely, L.~Williams, Attack surface definitions: A systematic literature review, Information and Software Technology 104 (2018) 94--103.

\bibitem{iorga2021yggdrasil}
D.~Iorga, D.-G. Corlatescu, O.~Grigorescu, C.~Sandescu, M.~Dascalu, R.~Rughinis, Yggdrasil—early detection of cybernetic vulnerabilities from twitter, in: 2021 23rd International Conference on Control Systems and Computer Science (CSCS), IEEE, 2021, pp. 463--468.

\bibitem{queiroz2019eavesdropping}
A.~L. Queiroz, S.~Mckeever, B.~Keegan, Eavesdropping hackers: Detecting software vulnerability communication on social media using text mining, in: The Fourth International Conference on Cyber-Technologies and Cyber-Systems, 2019, pp. 41--48.

\bibitem{baccar2021automated}
K.~Baccar, Automated mapping of cve vulnerabilties to mitre att\&ck framework, Ph.D. thesis, Tekup (2021).

\bibitem{dionisio2019cyberthreat}
N.~Dion{\'\i}sio, F.~Alves, P.~M. Ferreira, A.~Bessani, Cyberthreat detection from twitter using deep neural networks, in: 2019 international joint conference on neural networks (IJCNN), IEEE, 2019, pp. 1--8.

\bibitem{tang2023csgvd}
W.~Tang, M.~Tang, M.~Ban, Z.~Zhao, M.~Feng, Csgvd: A deep learning approach combining sequence and graph embedding for source code vulnerability detection, Journal of Systems and Software 199 (2023) 111623.

\bibitem{sun2023automatic}
X.~Sun, Z.~Ye, L.~Bo, X.~Wu, Y.~Wei, T.~Zhang, B.~Li, Automatic software vulnerability assessment by extracting vulnerability elements, Journal of Systems and Software 204 (2023) 111790.

\bibitem{CVSS}
NVD, Cvss, \url{https://nvd.nist.gov/vuln-metrics/cvss/v3-calculator}.

\bibitem{rahman2024attackers}
M.~R. Rahman, S.~K. Basak, R.~Mahdavi-Hezaveh, L.~A. Williams, Attackers reveal their arsenal: An investigation of adversarial techniques in cti reports, CoRR (2024).

\bibitem{son2023introduction}
S.~B. Son, S.~Park, H.~Lee, Y.~Kim, D.~Kim, J.~Kim, Introduction to mitre att\&ck: Concepts and use cases, in: 2023 International Conference on Information Networking (ICOIN), IEEE, 2023, pp. 158--161.

\bibitem{irshad2023cyber}
E.~Irshad, A.~B. Siddiqui, Cyber threat attribution using unstructured reports in cyber threat intelligence, Egyptian Informatics Journal 24~(1) (2023) 43--59.

\bibitem{MITRE}
MITRE, Mitre att\&ck, \url{https://attack.mitre.org/} (2024).

\bibitem{satvat2021extractor}
K.~Satvat, R.~Gjomemo, V.~Venkatakrishnan, Extractor: Extracting attack behavior from threat reports, in: 2021 IEEE European Symposium on Security and Privacy (EuroS\&P), IEEE, 2021, pp. 598--615.

\bibitem{Metasploit}
Metasploit, \url{https://www.metasploit.com/}.

\bibitem{Metasploitprocedure}
Fin6 group, \url{https://attack.mitre.org/groups/G0037/}.

\bibitem{wu2021price}
Y.~Wu, Q.~Liu, X.~Liao, S.~Ji, P.~Wang, X.~Wang, C.~Wu, Z.~Li, Price tag: towards semi-automatically discovery tactics, techniques and procedures of e-commerce cyber threat intelligence, IEEE Transactions on Dependable and Secure Computing (2021).

\bibitem{noor2019machine}
U.~Noor, Z.~Anwar, T.~Amjad, K.-K.~R. Choo, A machine learning-based fintech cyber threat attribution framework using high-level indicators of compromise, Future Generation Computer Systems 96 (2019) 227--242.

\bibitem{muennighoff2022mteb}
N.~Muennighoff, N.~Tazi, L.~Magne, N.~Reimers, Mteb: Massive text embedding benchmark, arXiv preprint arXiv:2210.07316 (2022).

\bibitem{colangelo2025comparative}
M.~T. Colangelo, M.~Meleti, S.~Guizzardi, E.~Calciolari, C.~Galli, A comparative analysis of sentence transformer models for automated journal recommendation using pubmed metadata, Big Data and Cognitive Computing 9~(3) (2025) 67.

\bibitem{choi2021evaluation}
H.~Choi, J.~Kim, S.~Joe, Y.~Gwon, Evaluation of bert and albert sentence embedding performance on downstream nlp tasks, in: 2020 25th International conference on pattern recognition (ICPR), IEEE, 2021, pp. 5482--5487.

\bibitem{pretrained}
Sentence transformers, accessed: May 2, 2024. \url{https://www.sbert.net/docs/pretrained_models.html} (2024).

\bibitem{wang2020minilm}
W.~Wang, F.~Wei, L.~Dong, H.~Bao, N.~Yang, M.~Zhou, Minilm: Deep self-attention distillation for task-agnostic compression of pre-trained transformers, Advances in Neural Information Processing Systems 33 (2020) 5776--5788.

\bibitem{HintonEtAl2015}
G.~Hinton, O.~Vinyals, J.~Dean, Distilling the knowledge in a neural network, stat 1050 (2015) 9.

\bibitem{okonkwo2023leveraging}
O.~Okonkwo, A.~Dridi, E.~Vakaj, Leveraging word embeddings and transformers to extract semantics from building regulations text, in: Proceedings of the 11th Linked Data in Architecture and Construction Workshop, 2023.

\bibitem{siino2024text}
M.~Siino, I.~Tinnirello, M.~La~Cascia, Is text preprocessing still worth the time? a comparative survey on the influence of popular preprocessing methods on transformers and traditional classifiers, Information Systems 121 (2024) 102342.

\bibitem{HinidumaEtAl2025}
K.~Hiniduma, S.~Byna, J.~L. Bez, \href{https://doi.org/10.1145/3722214}{Data readiness for ai: A 360-degree survey}, ACM Comput. Surv. 57~(9) (Apr. 2025).
\newblock \href {https://doi.org/10.1145/3722214} {\path{doi:10.1145/3722214}}.
\newline\urlprefix\url{https://doi.org/10.1145/3722214}

\bibitem{hyperparams}
Semantic textual similarity, \url{https://github.com/UKPLab/sentence-transformers/blob/master/examples/sentence_transformer/training/sts/README.md}.

\bibitem{lobo2022cost}
A.~Lobo, P.~Oliveira, P.~Sampaio, P.~Novais, Cost-sensitive learning and threshold-moving approach to improve industrial lots release process on imbalanced datasets, in: International Symposium on Distributed Computing and Artificial Intelligence, Springer, 2022, pp. 280--290.

\bibitem{sheng2006thresholding}
V.~S. Sheng, C.~X. Ling, Thresholding for making classifiers cost-sensitive, in: Aaai, Vol.~6, 2006, pp. 476--481.

\bibitem{NiEtAl2021}
J.~Ni, C.~Qu, J.~Lu, Z.~Dai, G.~H. Ábrego, J.~Ma, V.~Y. Zhao, Y.~Luan, K.~B. Hall, M.-W. Chang, Y.~Yang, \href{https://arxiv.org/abs/2112.07899}{Large dual encoders are generalizable retrievers} (2021).
\newblock \href {http://arxiv.org/abs/2112.07899} {\path{arXiv:2112.07899}}.
\newline\urlprefix\url{https://arxiv.org/abs/2112.07899}

\bibitem{sjoberg2022construct}
D.~I. Sj{\o}berg, G.~R. Bergersen, Construct validity in software engineering, IEEE Transactions on Software Engineering 49~(3) (2022) 1374--1396.

\bibitem{Shadish2002expdesign}
W.~R. Shadish, T.~D. Cook, D.~T. Campbell, Experimental and quasi-experimental designs for generalized causal inference / William R. Shedish, Thomas D. Cook, Donald T. Campbell, Wadsworth, Cengage Learning, Belmont, CA, 2002.

\bibitem{wohlin2012experimentation}
C.~Wohlin, P.~Runeson, M.~H{\"o}st, M.~C. Ohlsson, B.~Regnell, A.~Wessl{\'e}n, et~al., Experimentation in software engineering, Vol. 236, Springer, 2012.

\bibitem{exploitdb}
Offsec, Exploit-db, \url{https://www.exploit-db.com/}.

\bibitem{kuppa2021linking}
A.~Kuppa, L.~Aouad, N.-A. Le-Khac, Linking cve’s to mitre att\&ck techniques, in: Proceedings of the 16th International Conference on Availability, Reliability and Security, 2021, pp. 1--12.

\bibitem{devlin2018bert}
J.~Devlin, M.-W. Chang, K.~Lee, K.~Toutanova, Bert: Pre-training of deep bidirectional transformers for language understanding, in: Proceedings of the 2019 conference of the North American chapter of the association for computational linguistics: human language technologies, volume 1 (long and short papers), 2019, pp. 4171--4186.

\bibitem{sun2021generating}
J.~Sun, Z.~Xing, H.~Guo, D.~Ye, X.~Li, X.~Xu, L.~Zhu, Generating informative cve description from exploitdb posts by extractive summarization, CoRR (2021).

\bibitem{lakhdhar2021machine}
Y.~Lakhdhar, S.~Rekhis, Machine learning based approach for the automated mapping of discovered vulnerabilities to adversial tactics, in: 2021 IEEE Security and Privacy Workshops (SPW), IEEE, 2021, pp. 309--317.

\bibitem{grigorescu2022cve2att}
O.~Grigorescu, A.~Nica, M.~Dascalu, R.~Rughinis, Cve2att\&ck: Bert-based mapping of cves to mitre att\&ck techniques, Algorithms 15~(9) (2022) 314.

\bibitem{ampel2021linking}
B.~Ampel, S.~Samtani, S.~Ullman, H.~Chen, Linking common vulnerabilities and exposures to the mitre att\&ck framework: A self-distillation approach, 1st KDD Workshop on AI-enabled Cybersecurity Analytics (2021).

\bibitem{liu2019roberta}
Y.~Liu, M.~Ott, N.~Goyal, J.~Du, M.~Joshi, D.~Chen, O.~Levy, M.~Lewis, L.~Zettlemoyer, V.~Stoyanov, Roberta: A robustly optimized bert pretraining approach, arXiv preprint arXiv:1907.11692 (2019).

\bibitem{hemberg2022sourcing}
E.~Hemberg, A.~Srinivasan, N.~Rutar, U.-M. O’Reilly, Sourcing language models and text information for inferring cyber threat, vulnerability and mitigation relationships (2022).

\bibitem{kanakogi2021tracing}
K.~Kanakogi, H.~Washizaki, Y.~Fukazawa, S.~Ogata, T.~Okubo, T.~Kato, H.~Kanuka, A.~Hazeyama, N.~Yoshioka, Tracing cve vulnerability information to capec attack patterns using natural language processing techniques, Information 12~(8) (2021) 298.

\bibitem{othman2024comparison}
R.~Othman, B.~Rossi, B.~Russo, A comparison of vulnerability feature extraction methods from textual attack patterns, in: 2024 50th Euromicro Conference on Software Engineering and Advanced Applications (SEAA), IEEE, 2024, pp. 419--422.

\bibitem{hemberg2021linking}
E.~Hemberg, J.~Kelly, M.~Shlapentokh-Rothman, B.~Reinstadler, K.~Xu, N.~Rutar, U.-M. O'Reilly, Linking threat tactics, techniques, and patterns with defensive weaknesses, vulnerabilities and affected platform configurations for cyber hunting (2021).
\newblock \href {http://arxiv.org/abs/2010.00533} {\path{arXiv:2010.00533}}.

\bibitem{ramos2003using}
J.~Ramos, et~al., Using tf-idf to determine word relevance in document queries, in: Proceedings of the first instructional conference on machine learning, Vol. 242, Citeseer, 2003, pp. 29--48.

\bibitem{lau2016empirical}
J.~H. Lau, T.~Baldwin, An empirical evaluation of doc2vec with practical insights into document embedding generation, in: Proceedings of the 1st Workshop on Representation Learning for NLP, Association for Computational Linguistics, 2016.

\bibitem{kanakogi2022comparative}
K.~Kanakogi, H.~Washizaki, Y.~Fukazawa, S.~Ogata, T.~Okubo, T.~Kato, H.~Kanuka, A.~Hazeyama, N.~Yoshioka, Comparative evaluation of nlp-based approaches for linking capec attack patterns from cve vulnerability information, Applied Sciences 12~(7) (2022) 3400.

\bibitem{SecurityWeek}
SecurityWeek, News vulnerabilities, \url{https://www.securityweek.com/category/vulnerabilities/}.

\end{thebibliography}

% \bibliographystyle{cas-model2-names}

% \bibliography{cas-refs}

\noindent
\begin{minipage}{0.18\textwidth}
   \centering
   \includegraphics[width=\textwidth]{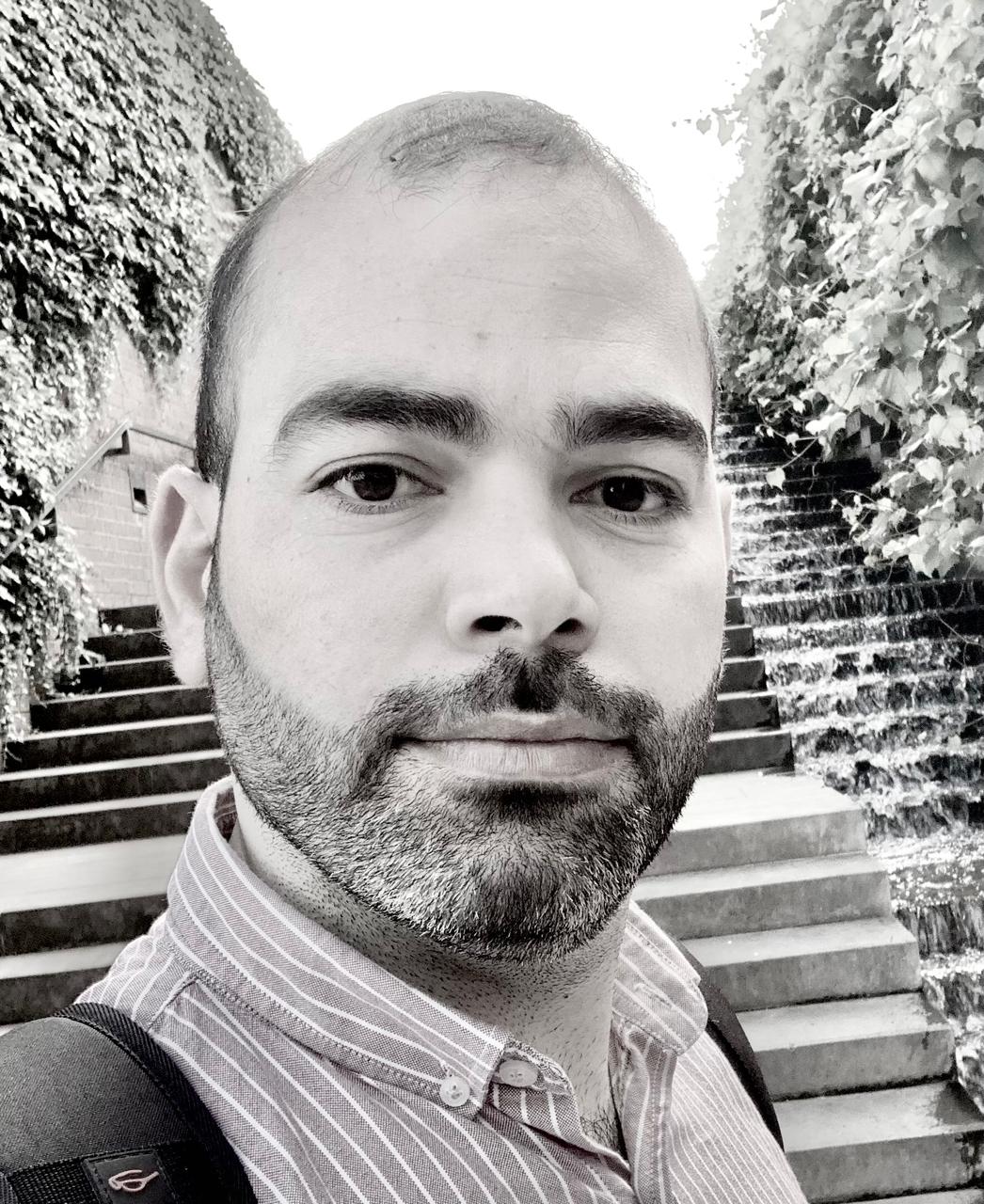}
\end{minipage}%
\hspace{0.02\textwidth}%
\begin{minipage}{0.78\textwidth}
      Refat Othman is a PhD candidate in Advanced-Systems Engineering at the Free University of Bozen-Bolzano, specializing in cybersecurity. He works as a researcher at the Cybersecurity Laboratory (CSLab), where his research focuses on automating vulnerability detection from textual cyberattack descriptions using sentence transformer models, while leveraging datasets from MITRE repositories. Refat holds both a master’s and bachelor’s degree in Software Engineering and Computer Science from Birzeit University, where he also served as an instructor. His industry experience includes roles as a QA Team Leader and Senior Software Engineer at NVIDIA and Asal Technologies.
   \end{minipage}
\vspace{0.3em} % small manual spacing if needed

   \noindent
\begin{minipage}{0.18\textwidth}
   \centering
   \includegraphics[width=\textwidth]{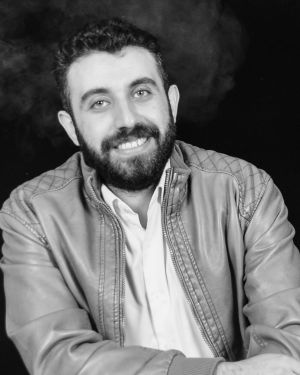}
   \end{minipage}%
\hspace{0.02\textwidth}%
\begin{minipage}{0.78\textwidth}
      Diaeddin Rimawi received his Ph.D. in Advanced-Systems Engineering from the Free University of Bozen-Bolzano, where he introduced the concept of green resilience in AI-enabled cyber-physical systems. Before his Ph.D., he worked in both academia and industry as a university instructor, head of instructors at a coding bootcamp, and software engineer. He is currently a Cybersecurity Technologist at the CSLab of the Free University of Bozen-Bolzano. His research interests include optimization, game theory, and reinforcement learning to support sustainable and resilient decision-making under unforeseen disruptions and adversarial attacks.
   \end{minipage}
\vspace{0.3em} % small manual spacing if needed

\noindent
\begin{minipage}{0.18\textwidth}
   \centering
   \includegraphics[width=\textwidth]{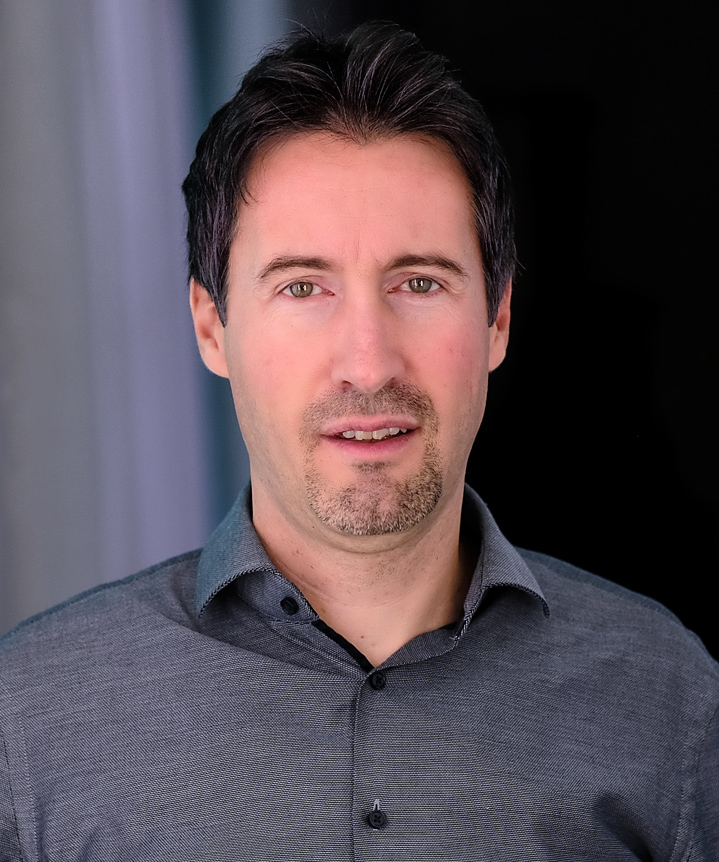}
   \end{minipage}%
\hspace{0.02\textwidth}%
\begin{minipage}{0.78\textwidth}
      Bruno Rossi is Associate Professor at the Lab of Software Architectures and Information Systems, Faculty of Informatics, Masaryk University, Brno, Czech Republic. In 2008 he received the Ph.D. degree in Computer Science from the Free University of Bozen-Bolzano, Bolzano, Italy. He is part of the C4e project, “Center of Excellence for Cybercrime, Cybersecurity and Protection of Critical Information Infrastructures,” and has taken part to key European (COSPA, STREP FP6) and Italian projects (ArtDeco, FIRB 36 months). He is involved as an Active Member in several journal and conference program committees in software engineering. His research interests include empirical software engineering research and cyberphysical systems, with specific focus on smart grids.
   \end{minipage}

\vspace{0.8em} % small manual spacing if needed
 
\noindent
% \begin{minipage}{0.18\textwidth}
%    \centering
%    \includegraphics[width=\textwidth]{Fig/Barbaranew.jpg}
%    \end{minipage}%
   \begin{minipage}{0.18\textwidth}
   \centering
   \includegraphics[width=\textwidth]{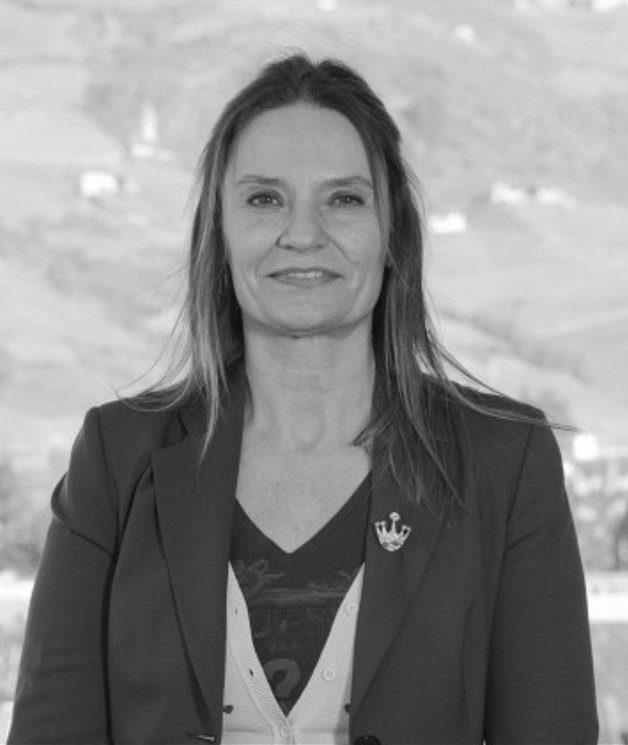}
   \end{minipage}%
   % \begin{minipage}{0.18\textwidth}
   % \centering
   % \includegraphics[width=\textwidth,height=1.18\textwidth]{Fig/BarbaraN1.jpg}
   % \end{minipage}%
\hspace{0.02\textwidth}%
\begin{minipage}{0.78\textwidth}
Barbara Russo is a Full Professor of Computer Science at the Faculty of Engineering at the Free University of Bozen-Bolzano. She has served as Vice Dean for Research at the Faculty of Computer Science and Technologies and has coordinated scientific programs for international symposia and doctoral schools. Barbara Russo is an Associate Editor for the International Journal of Information and Software Technology (Elsevier) and acts as a reviewer for several high-level journals and conferences in software engineering e.g., ICSE, MSR, EMSE, FSE) journals (TSE, TOSEM, ESE, ASE) in Software Engineering. She founded and has been organizing the international doctoral school in software engineering in collaboration with the University of Innsbruck for the past 10 years, with the 2023 edition focusing on cybersecurity. She has published 150 articles in international mathematics and computer science journals, which have received over 3500 citations (h index=29). Her expertise lies in software development and maintenance aimed at ensuring systems with high standards of reliability, scalability, performance, and security. In the last four years, prof. Russo’ research specializes in data extraction using Deep Learning techniques to reconstruct system and user behaviour and detect anomalies and code vulnerabilities.

Since 2023 she is coordinating the Cyber Security Lab of the Free University of Bolzano, supported by the FESR 2021-2027 program.
Since 2024 she is coordinating the PhD program in Advanced Systems Engineering of the Faculty of Engineering of the Free University of Bozen-Bolzano, Italy.
In 2020-2023, she was the vice-dean for research at the Faculty of Computer Science of the Free University of Bozen-Bolzano, Italy Since 2022 she is unit coordinator of BeT project funded by PRIN ministerial fund and she was unit coordinator of other two PRIN projects (Idea and GAUSS).
In 2014-2019 she was project coordinator of the Erasmus Mundus in Software Engineering of the Faculty of computer Science of the Free University of Bozen-Bolzano, Italy (successfully audited by the European Commission in 2022) (2M EURO).
   \end{minipage}
% \end{multicols}
\end{document}